\documentclass[aps,prd,preprint,superscriptaddress,floatfix,showpacs,showkeys]{revtex4-1}
\usepackage{epsfig}
\usepackage{dcolumn}
\usepackage{bm}
\usepackage[latin5]{inputenc}
\usepackage{graphics}
\usepackage{graphicx}
\usepackage{epsfig}
\usepackage{amssymb}
\usepackage{amsmath}
\usepackage{hyperref}
\begin{document}
\title[Thermodynamic Comparison of the GSWSPE ]{A Comparative Interpretation of the Thermodynamic Functions of a Relativistic Bound State problem proposed with an Attractive or a Repulsive Surface Effect}

\author{B.C. L\"{u}tf\"{u}o\u{g}lu}
\affiliation{Department of Physics, Faculty of Science, Akdeniz University, 07058
Antalya, Turkey}

\affiliation{Department of Physics, Faculty of Science, University of Hradec Kr\'{a}lov\'{e}, Rokitansk\'{e}ho 62, 500\,03 Hradec Kr\'{a}lov\'{e}, Czechia}

\author{J. K\v{r}\'{i}\v{z}}
\affiliation{Department of Physics, Faculty of Science, University of Hradec Kr\'{a}lov\'{e}, Rokitansk\'{e}ho 62, 500\,03 Hradec Kr\'{a}lov\'{e}, Czechia}
\date{\today}
\begin{abstract}
Generalized symmetric Woods-Saxon potential (GSWSP) energy well has a significant importance not only in nuclear physics, but also in atomic and molecular physics. A GSWSP energy well takes the surface effects (describing e.g. repulsive interaction at the nucleus edge in nuclear physics) into account in addition to the volume effect. These effects can be, in general, both repulsive or attractive. In this paper, the recently obtained bound state solution of the Klein-Gordon equation with vector and scalar GSWSP energy in the spin symmetry limit is used to calculate a neutral pion's energy spectra in attractive and repulsive cases via various potential parameters. Then, the spectra are employed to find the thermodynamic functions such as Helmholtz free energy, entropy, internal energy, and specific heat. These functions in the attractive and repulsive cases are compared comprehensively. Finally, the role of the shape parameters on the thermodynamic functions in repulsive and attractive cases, respectively, is analyzed.
\end{abstract}
\maketitle

\section{Introduction}\label{sec:Intro}

In 1954, Woods and Saxon had proposed a continuous potential energy with three parameters \cite{Woods_et_al_1954} to replace the square well potential energy in the optical model description of the scattering of a nuclear particle by a nucleus \cite{Feshbach_1958}. Since then, the Woods-Saxon potential (WSP) energy is used in huge amount of applications in various branches of physics. For instance, in nuclear physics some of recent studies can be given in \cite{esbensenDavids2000, michelNazarewiczPloszajczakBennaceur2002, michelNazarewiczPloszajczak2004, Volya_et_al_2006, Boztosun_et_al_2008, Aygun_et_al_2010, coban2012, ikhdair2013, Zhang_et_al_2014, soylu2015, salamon2016, Jaganathen_et_al_2017, Betan_et_al_2018} in addition to two reports \cite{satchler1991, brandanSatchler1997}, in atomic and molecular physics \cite{Leforestier_et_al_1983, Seideman_1992, Seideman_et_al_1992, Costa_et_al_1999}, in non relativistic \cite{Book_Flugge, Saha_et_al_2011, Feizi_et_al_2011, Pahlavani_et_al_2012, Nikram_et_al_2016} and in relativistic quantum mechanics \cite{Kennedy_2002, Guo_et_al_2002, Rojas_et_al_2005, Ikhdair_et_al_2007, Arda_et_al_2008, Hassanabadi_et_al_2013, Olgar_et_al_2015, Capak_et_al_2016}.

Satchler suggested that the surface effects at the nuclear edge should also be taken into account in WSP energy. Therefore, he proposed the generalized WSP (GWSP) energy by offering an additional term to the WSP energy \cite{Book_Satchler}.  We introduce this additional term and describe its physical meaning in section \ref{sec:poten}.  GWSP energy is examined in various papers as well \cite{Kobos_et_al_1984, Koura_et_al_2000, Boztosun_et_al_2005, Badalov_et_al_2009, Kocak_et_al_2010, Dapo_et_al_2012, Candemir_et_al_2014, Bayrak_et_al_2015, Lutfuoglu_et_al_2016, Liendo_et_al_2016, Lutfuoglu_ctp_2018, Lutfuoglu_et_al_2018_133, Lutfuoglu_cjp_2018, Lutfuoglu_2018, Lutfuoglu_et_al_2018_Ikot}.

On the other hand, in statistical mechanics, the thermodynamic behavior of a macroscopic system is being described by the laws which govern the behavior of the microscopic elements such as atoms, molecules, dipole and magnetic moments \cite{Book_Huang}. There, a very powerful fundamental tool, the partition function, is being used. Note that, if the exact analytical method \cite{Book_Reif} cannot be used to obtain a partition function, several approximative methods are  employed such that Wigner-Kirkwood expansion method \cite{Wigner_1932, Kirkwood_1933}, semi-classical expansion method \cite{Korsch_1979}, Pad\'e approximant method \cite{Gibson_1984}, etc.

In the last decade, the investigations on the thermodynamic functions became popular. Dong \emph{et al.} used symmetries to explore the non relativistic exact solution of the one dimensional system which is governed by a harmonic oscillator and an inverse square potential. There, they obtained the vibrational partition functions, thus, the vibrational thermodynamic functions. They showed that the parameter controlling the depth of the well does not contribute on the  entropy and specific heat functions \cite{Dong_et_al_2007}. Pacheco \emph{et al.} considered  one and three dimensional Dirac oscillators in a thermal bath and determined their canonical partition functions. Then, they analyzed the main thermodynamic functions and concluded that in higher values of temperature the specific heat functions are much greater \cite{Pacheco_et_al_2003, Pacheco_et_al_2014}. Note that, in 2013, Franco-Villafan\~{n}e \emph{et al.} succeeded the first experimental realization on the one dimensional Dirac oscillator \cite{Villafane_et_al_2013}. Hassanabadi \emph{et al.} solved the Dirac equation in $3+1$ dimension with a non minimal coupling of a static electromagnetic potential that is constituted with both the Aharonov-Bohm and magnetic monopole fields. Then, they obtained the partition function and examined the thermodynamic properties such as  Helmholtz free energy, entropy, mean energy and the specific heat \cite{Hassanabadi_et_al_2015}.

The well-known potential energies used in nuclear, atomic and molecular physics are being a subject of the investigation of thermodynamic functions, too. Ikhdair \emph{et al.} used the non-relativistic and relativistic solutions of the P\"oschl-Teller potential to examine the thermodynamic functions \cite{Ikhdair_et_al_2013}. Onate \emph{et al.} employed the generalized  P\"oschl-Teller potential with hyperbolical potential energy to obtain some thermodynamic properties \cite{Onate_et_al_2015}. Ikot \emph{et al.} investigated exponential-type molecule potentials to investigate the thermodynamic functions in D dimensions \cite{Ikot_et_al_BCL_2016}. Onyeaju \emph{et al.} obtained the thermodynamic functions of a confined particle in  the Dirac well of a deformed Hylleraas plus deformed Woods-Saxon potential energy \cite{Onyeaju_et_al_2017}. Valencia-Ortega \emph{et al.} examined the diatomic molecule's thermodynamic properties under $SO(2,1)$ anharmonic Eckart potential energy \cite{Ortega_et_al_2018}. In a very recent paper, Okorie \emph{et al.} examined the modified Mobius square potential energy in the bound state solutions of Schr\"odinger equation to obtain the thermodynamic properties \cite{Okorie_et_al_2018}.

In 1932, Rosen-Morse potential (RMP) energy was proposed by their authors, for the investigation of the vibrational states of polyatomic molecules such as NH$_3$ molecule \cite{Rosen_et_al_1932}. Ocak \emph{et al.} used the modified RMP to describe the internal vibration of the sodium dimer to calculate the thermodynamic quantities for the Na$_2$ $(5^1\Delta_g)$  molecule \cite{Ocak_et_al_2018}. The improved RMP, from now on (IRMP), is employed to calculate the  vibrational energies of the Cs$_2$, Na$_2$, Li$_2$ and gaseous BBr  molecules \cite{Chen_et_al_2013, Hu_et_al_2014, Jia_et_al_2017_667_211, Wang_et_al_2017}. In their recent works Jia \textit{et al.} examined the IRMP energy to predict the molar entropy  and enthalpy values in addition to the  Gibbs free energies for the NO and P$_2$ molecules in a wide temperature range  \cite{Jia_et_al2018a, Jia_et_al2018b, Peng_et_al_2018}. In 2012, Jia \emph{et al.} announced the equivalency of the IRMP to the  improved Tietz potential (ITP) \cite{Jia_et_al_2012}. Not long ago, Jia  \emph{et al.} used the Poisson summation formula to calculate the vibrational partition function of the ITP  \cite{Jia_et_al_2017_676}. Very recently, ITP energy is employed to describe the internal vibration of a molecule via ITP energy to investigate the entropy and specific heat for some gaseous substances \cite{Jia_et_al_2018_692, Khordad_et_al_2019}.

In 2012,  Zhang \textit{et al.} derived a very crucial relation in between the IRMP and GWSP energies \cite{Zhang_et_al_2012}.  They proved that in diatomic molecules the IRMP and GWSP energies are the same empirical potential energies.  Therefore, the GWSP energy turns to be a good candidate for being used in molecular physics problems.

One of the authors of the present paper, (BCL), with his collaborators examined the symmetric form of the GWSP, namely generalized symmetric Woods-Saxon potential (GSWSP), energy in the Schr\"odinger equation in one dimension \cite{Lutfuoglu_et_al_2016}. They obtained two transcendental equations to calculate the energy spectrum.  Shortly afterward, he with another collaborator used those equations  to discuss the thermodynamic properties of a bound nucleon in a light nucleus \cite{Lutfuoglu_et_al_2016_17}. In the next work, they considered an alpha particle that is confined in a heavy nucleus. They analyzed the  thermodynamic functions of the alpha particle \cite{Lutfuoglu_et_al_2017_21}. Note that, within both studies, surface effects were taken to be only repulsive. Therefore, the bound state energy spectrum was allowed to be divided into two subsets, namely tight-bound spectrum and quasi-bound spectrum. The discussion on the thermodynamic functions was executed via excluding and including the quasi-bound energy spectrum to the tight-bound spectrum.

We examined the GSWSP energy in a relativistic equation, namely in the Klein-Gordon (KG) equation. First, we explored the scattering solution in the KG equation in the limits of the spin symmetry (SS) and pseudospin symmetry (PSS) \cite{Lutfuoglu_et_al_2018_133}.  Then, BCL analyzed the bound state solution in the same limits, he observed a very surprising result, such that only in the SS limit a bound state spectrum could have existed \cite{Lutfuoglu_2018}. He compared the WSP with GSWSP energies in the context of statistical mechanics first in the non-relativistic \cite{Lutfuoglu_ctp_2018} and then in the relativistic approaches \cite{Lutfuoglu_cjp_2018}.

GSWSP energy has a flexible structure since the surface effects can be suggested to be repulsive or attractive. Therefore, it easily can be employed within four different physical problems. For instance, when a GSWSP energy barrier is investigated, an extra barrier or a pocket near the surface can be considered by employing a repulsive or attractive surface effect, respectively. Alike the barrier problems, when a GSWSP well is considered, a barrier or an additional pocket near the surface can be obtained with a repulsive or an attractive effect, respectively. In the literature, up to our knowledge, all discussions on the examining the thermodynamic functions were executed by the use of a repulsive surface effect only. Therefore, in this paper, our main motivation is to obtain and compare the thermodynamic functions of a confined particle in the GSWSP energy wells that have attractive or repulsive surface effects. In addition to the ongoing discussion,  we investigate how the shape parameters, which adjust the slope and the effective length of the well, are correlated with the thermodynamic functions in the existence of either repulsive or attractive surface effects. The obtained results can bring a new interpretation of the thermodynamics of the diatomic molecules as the consequence of Zhang \emph{et al.}'s paper. Moreover, the missing bricks on the literature wall are expected to be filled.

The structure of the paper is as follows. In section \ref{sec:poten}, we introduce the GSWSP energy and then we demonstrate the comparisons of the potential energies used. In section \ref{sec:KG_Bound_State}, we present a very brief solution of the KG equation that was obtained in our previous paper \cite{Lutfuoglu_2018}. We divide the section \ref{sec:EnergySpectra} into two subsections. We calculate and then tabulate the energy spectra in the presence of the repulsive or attractive surface effects in subsections \ref{subsec:Repulsive} and \ref{subsec:Attractive}, respectively. We introduce the thermodynamic functions that we examine in section \ref{sec:Thermo}. Then, we investigate those thermodynamic functions and compare the results within eighteen graphs in section \ref{sec:Discuss}. Finally, in section \ref{sec:Concl} we conclude the paper.

\section{Generalized Symmetric Woods-Saxon Potential Energy}\label{sec:poten}
We consider the GSWSP energy well with the following form
\begin{eqnarray}
  V(x)\equiv\theta{(-x)}\Bigg[\frac{-V_0}{1+e^{-a(x+L)}}+ \frac{W_0e^{-a(x+L)}}{\big(1+e^{-a(x+L)}\big)^2}\Bigg]+ \theta{(x)}\Bigg[\frac{-V_0}{1+e^{a(x-L)}}+  \frac{W_0e^{a(x-L)}}{\big(1+e^{a(x-L)}\big)^2}\Bigg], \label{GSWSP}
 \end{eqnarray}
where $\theta$ stands for the Heaviside function. Three parameters $V_0$, $a$ and $L$ are common with WSP energy and positively defined. $V_0$  has a dimension of energy and represents potential energy well depth via the volume effects. $a$ is a measure of the slope of the well, and $L$ indicates the effective length.   $W_0$ determines the strength of the surface interaction effects and has a dimension of energy. Basically, it is derived from the spatial derivative of the volume effect, therefore it is linearly proportional with the three parameters of the WSP energy in addition to an anonymous parameter. This anonymous parameter can be negative or positive, and so as $W_0$ can be negative or positive. Therefore, attractive or repulsive surface interactions, depending on the physical problem, can be investigated. Note that, when the effective length is zero, the GSWSP reduces to the RMP \cite{Bayrak_et_al_2015}. Moreover, when the slope parameter is zero, a square well potential energy  is obtained from the GSWSP energy well \cite{Book_Flugge}.

In our previous paper \cite{Lutfuoglu_2018}, we found that  a confined particle's spectrum possesses only positive values if $V_0\leq \frac{m_oc^2}{2}$ in the presence of a scalar and vector potential energy in the SS limit. Note that, this resulting condition does not depend on whether the surface interactions are attractive or repulsive. Moreover, if $V_0$ has a value that is equal to the rest mass energy of the confined particle, then the spectrum's eigenvalues constitute from the full range of the KG interval. Note that, Klein paradox constrains the depth parameter with an upper limit value in the order of the $m_oc^2$ \cite{Calogeracos_et_al_1999}. Therefore, in this paper, we decide to assign two critical values, $\frac{m_oc^2}{2}$ and  $m_oc^2$, to the depth parameter.

Another discussion that we made in the previous article \cite{Lutfuoglu_2018}, was on the appearance of  a \emph{"barrier"} for the repulsive and a \emph{"pocket"} for the attractive surface effects. There, we showed such effects appear  if and only if $|W_0| > V_0$ condition is satisfied. Furthermore, the occurred barrier's height  in the repulsive case and pocket's depth in the attractive case can be calculated with equation $\frac{(V_0-W_0)^2}{4W_0}$. Therefore, we decide to assign the values $-2m_0c^2$ in the attractive and $2m_0c^2$ in the repulsive cases to the parameter $W_0$.

Moreoever, we observed that the measure of the slope parameter and the effective length parameter should satisfy $e^{a L} >>1$ condition in order to examine a smooth GSWSP energy well \cite{Lutfuoglu_2018}. Therefore, in this paper, we adjust the slope parameter to be $1$ $fm^{-1}$ and the effective length parameter to be $6$ $fm$. On the other hand to investigate their effects on the thermodynamic functions we have to assign new values compatible with the condition given above. Thus, we set the values $1.5$ $fm^{-1}$, $2$ $fm^{-1}$ to the parameter $a$ and $7$ $fm$, $8$ $fm$ to the parameter $L$.

\begin{figure}[!htb]
\centering
\includegraphics[totalheight=0.45\textheight]{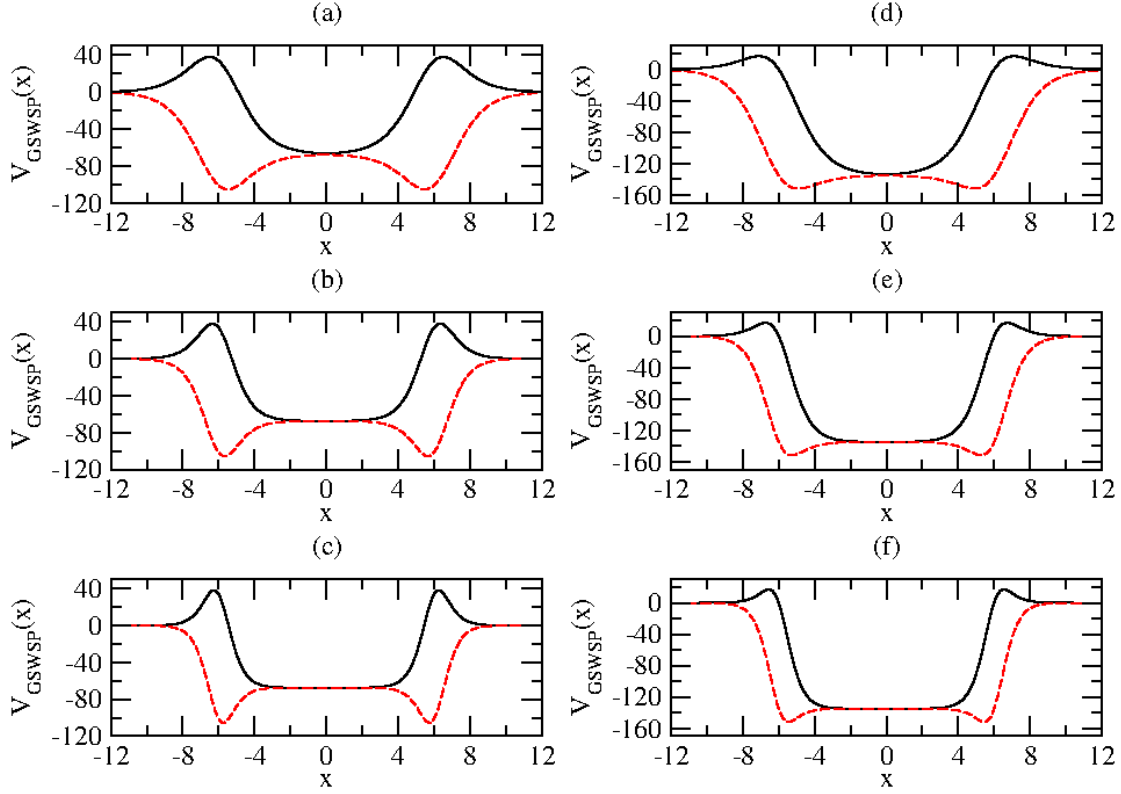}
   \caption{The examined GSWSP energy wells that have either repulsive (solid-black lines) or attractive (dashed-red lines) surface interactions are represented. The first and the second column has a potential depth parameter $V_0$, that is equal to $\frac{m_0c^2}{2}$ and $m_0c^2$,  respectively. The parameter of the strength of the surface interactions $W_0$ is equal to, $2m_0c^2$ in the repulsive, and $-2m_0c^2$  in the  attractive cases. The slope parameter has a different value in each row, and is equal to   $1$  $fm^{-1}$, $1.5$  $fm^{-1}$ and  $2$  $fm^{-1}$, respectively. The effective well distance parameter $L$ is chosen to be $6$  $ fm$ in all graphs. Note that all vertical axes are in units of $MeV$, while the distances are  $fm$.} \label{potfig01}
\end{figure}

\begin{figure}[!htb]
\centering
\includegraphics[totalheight=0.45\textheight]{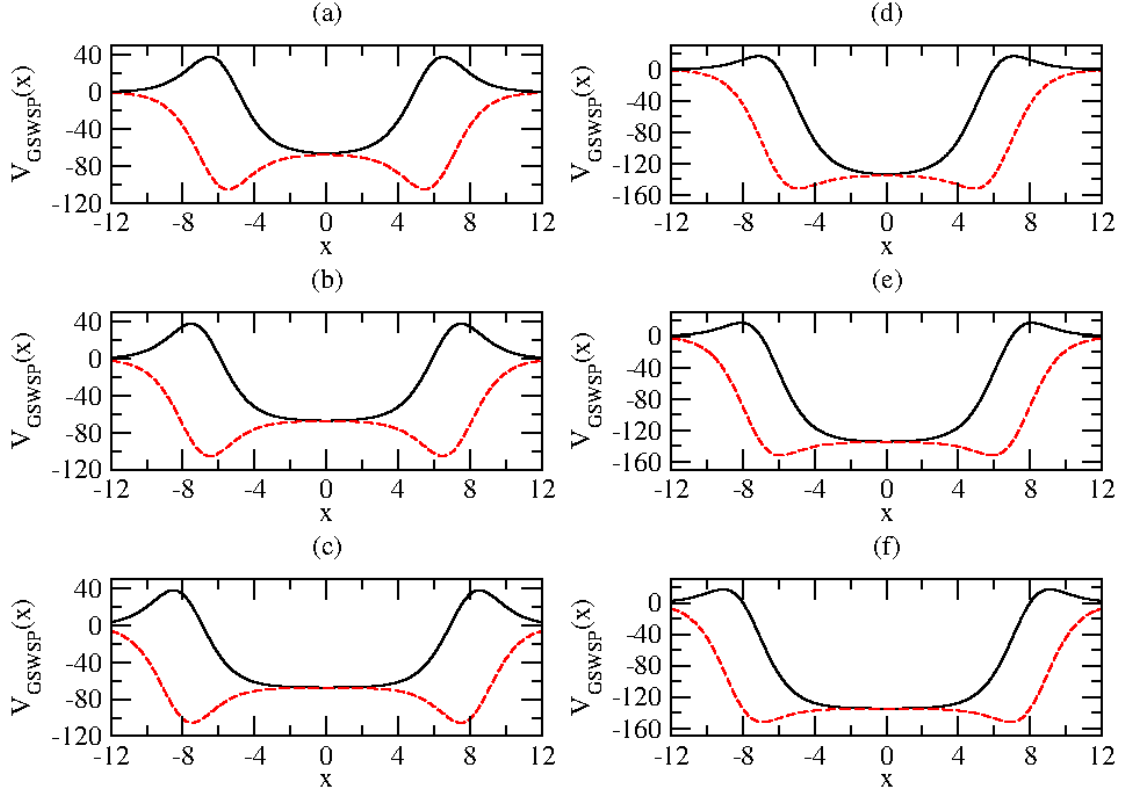}
   \caption{The examined GSWSP energy wells that have either repulsive (solid-black lines) or attractive (dashed-red lines) surface interactions are represented. The first and the second column has a potential depth parameter $V_0$, that is equal to $\frac{m_0c^2}{2}$ and $m_0c^2$,  respectively. The parameter of the strength of the surface interactions $W_0$ is equal to $2m_0c^2$  and $-2m_0c^2$ in the repulsive and attractive cases. The effective well distance parameter $L$ has a different value in each row, and is equal to   $6$  $fm$, $7$  $fm$ and  $8$  $fm$, respectively. The slope parameter $a$ is chosen to be $1$  $fm^{-1}$ in all graphs. Note that all vertical axes are in units of $MeV$, while the distances are  $fm$.} \label{potfig02}
\end{figure}

We summarize all the necessary GSWSP energy wells in fig. \ref{potfig01} and fig. \ref{potfig02}. In the both figures, in the first column we indicate the GSWSP energies that have depth parameter $\frac{m_0c^2}{2}$ whereas in the second column $m_0c^2$. All solid lines (black in colors) present the repulsive cases, while the dashed lines (red in colors) the attractive cases. In fig.\ref{potfig01}, in each row, we demonstrate the GSWSP energy wells with a new value of the slope parameter. Alike, in fig.\ref{potfig02}, we plot the GSWSP energy wells with a new value of the effective length parameter within each row.

\newpage

\begin{figure}[!htb]
\centering
\includegraphics[totalheight=0.45\textheight]{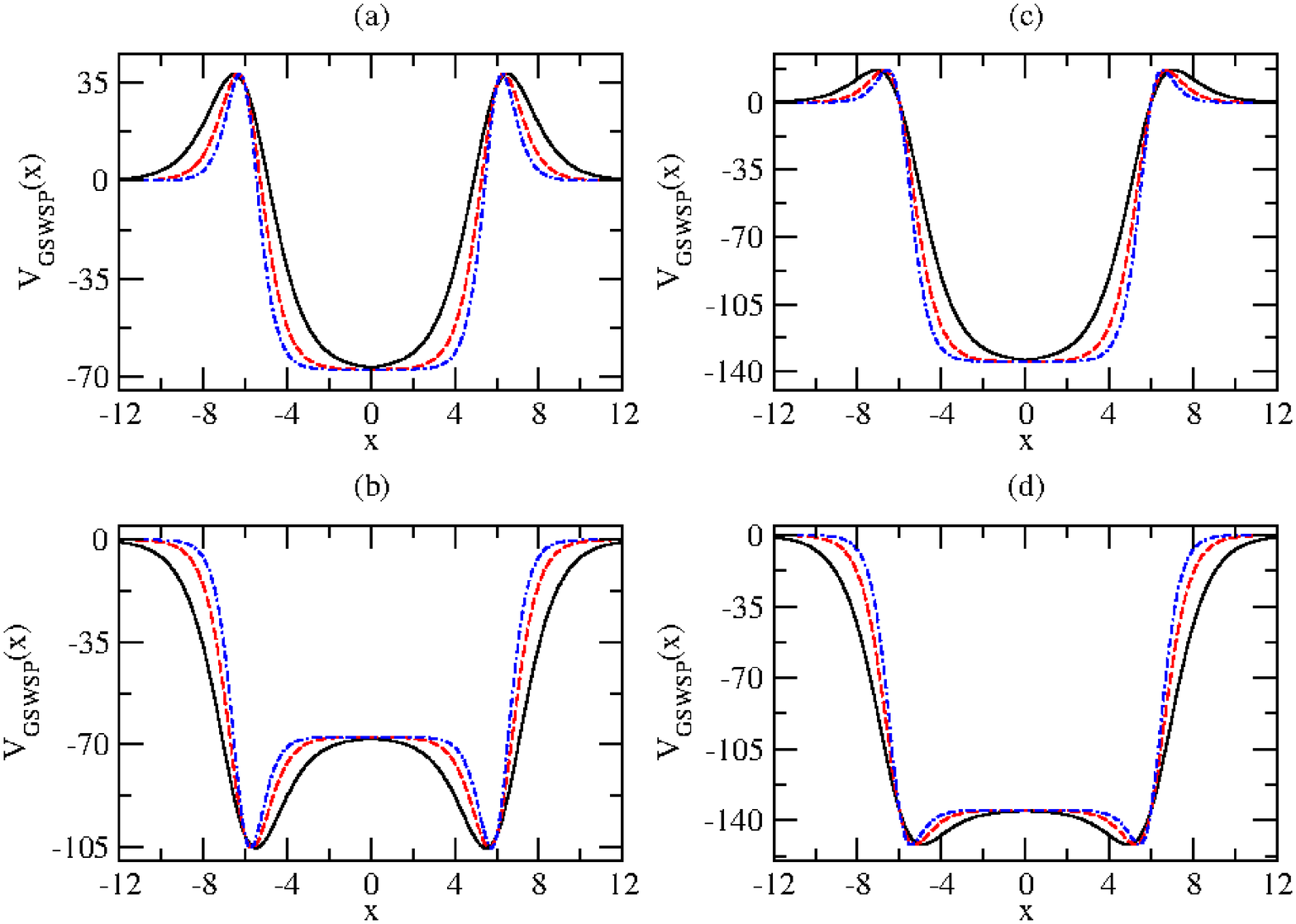}
   \caption{The dependence of the GSWSP energy wells on the slope parameters. The values of the slope parameter are given as follows: $a=1$  $fm^{-1}$ in the solid-black line, $a=1.5$  $fm^{-1}$ in the dashed-red line and $a=2$  $fm^{-1}$ in the dashed and dotted-blue line. The parameter $L$ is equal to $6$ $fm$ in every graph. The other parameters:  $V_0$ determines the depth of the wells is equal to  $\frac{m_0c^2}{2}$ in the first column and  $m_0c^2$ in the second column. Finally, in the  repulsive case $W_0=2m_0c^2$,  and in the attractive case $W_0=-2m_0c^2$. Note that all vertical axes are in units of $MeV$, while the distances are  $fm$.} \label{potfig03}
\end{figure}

\begin{figure}[!htb]
\centering
\includegraphics[totalheight=0.45\textheight]{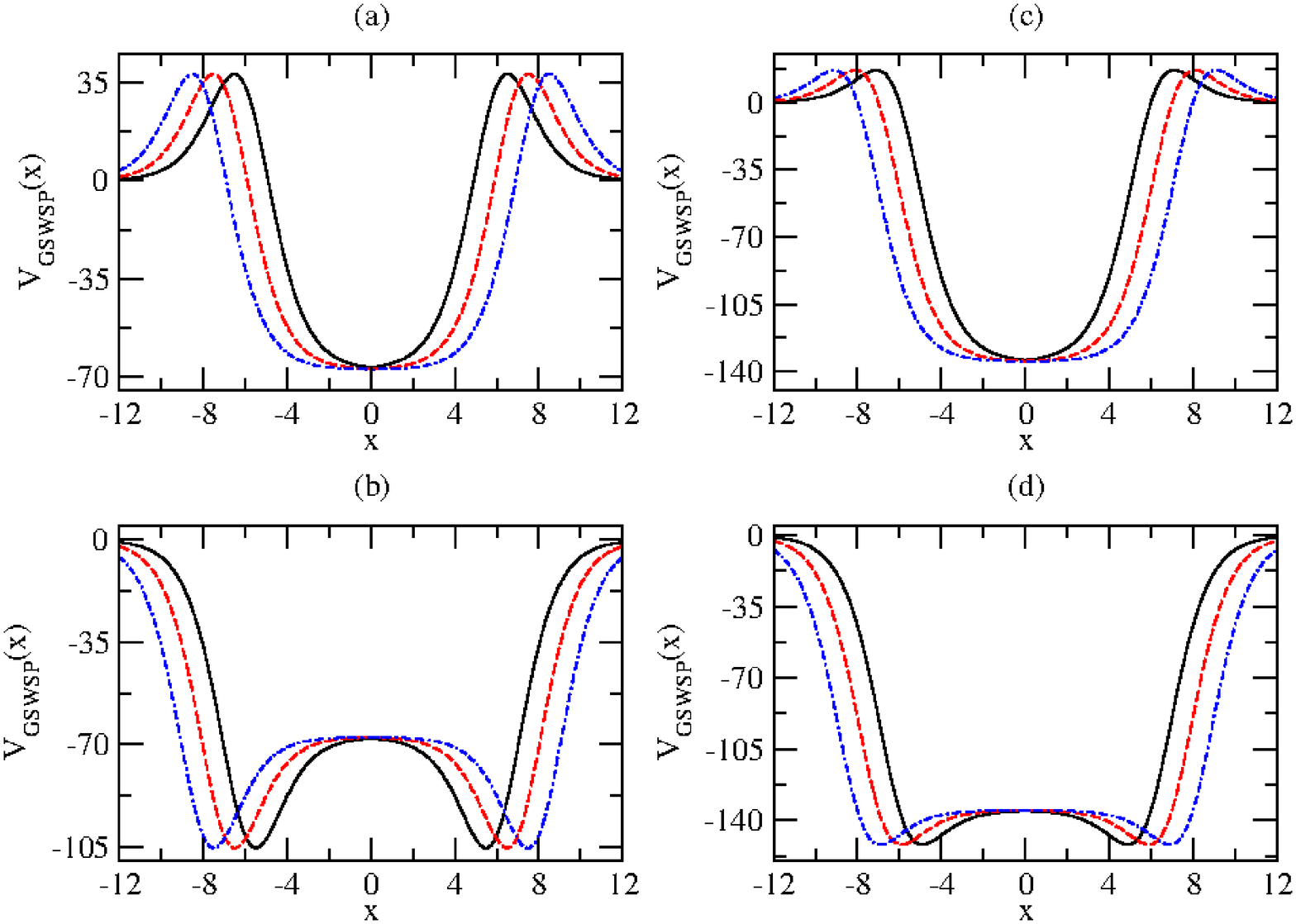}
   \caption{The dependence of the GSWSP energy wells on the effective length parameters. The values of the parameter $L$ are given as follows: $L=6$  $fm$ in the solid-black line, $L=7$  $fm$  in the dashed-red line and $L=7$  $fm$ in the dashed and dotted-blue line. The slope parameter $a$ is equal to $a=1$  $fm^{-1}$ in every graph. The other parameters:  $V_0$ determines the depth of the wells is equal to  $\frac{m_0c^2}{2}$ in the first column and  $m_0c^2$ in the second column. Finally, in the  repulsive case $W_0=2m_0c^2$,  and in the attractive case $W_0=-2m_0c^2$. Note that all vertical axes are in units of $MeV$, while the distances are  $fm$.} \label{potfig04}
\end{figure}

We present auxiliary plots in fig. \ref{potfig03}, and fig. \ref{potfig04}. Here, we intend to show the effects of the changes of the slope parameter and effective length parameter in each type of potential well within the subfigures of fig. \ref{potfig03} and fig. \ref{potfig04}, respectively.

\newpage

\section{The Klein-Gordon equation's bound state solution }\label{sec:KG_Bound_State}
KG equation describes the massive spin-$0$ relativistic particle's dynamics. There are different ways to take into account the interactions of these spinless particles. One of them, called minimal coupling, is the addition of a four-vector operator to the four-momentum operator. In the literature,  many successful applications are observed \cite{Greinerbook}. In a non-minimal way,  one can couple a scalar potential energy to the rest mass energy in addition to a four-vector coupling. Here, we introduce such a four-vector whose time-component differ from zero while spatial components not. We call this term as the "vector potential" and distinguish it from the "scalar potential" which is coupled to the rest mass energy. We assume the potential energies are time independent, thus the wave function is separable into time and spatial parts. Then, the KG equation becomes in one spatial dimension as follows
 \begin{eqnarray}
   \Bigg[\frac{d^2 }{d x^2}+\frac{1}{\hbar^2c^2} \bigg[\Big(E-V_v(x)\Big)^2 -\Big(m_0c^2+gV_s(x)\Big)^2\bigg] \Bigg]\phi(x) &=& 0.  \,\,\,\,\,\,\,\,\,\,\,\,\,\,\,\, \label{KGSS}
\end{eqnarray}
Here,  $\hbar$, $c$ and $m_0$  stand for the Planck constant, the speed of light in vacuum and the confined particle's rest mass, respectively. In the SS limit and in the strong regime, we use the well-known relation between the finite ranged potential energies, $V(x)\equiv V_v(x)=gV_s(x)$.  Then, we obtain
\begin{eqnarray}
   \Bigg[\frac{d^2 }{d x^2}+\frac{1}{\hbar^2c^2} \bigg[\Big(E^2-m_0^2c^4\Big) -2V(x)\Big(E+m_0c^2\Big)\bigg] \Bigg]\phi(x) &=& 0.  \,\,\,\,\,\,\,\,\,\,\,\,\,\,\,\, \label{KGS1}
\end{eqnarray}
In our previous article \cite{Lutfuoglu_2018}, we exerted the detailed calculations on the solution of the Eq. (\ref{KGS1}). Therefore, we avoid to give an unnecessary repetition, instead we present a very brief summary of the bound state solution. We use the  $x\rightarrow -x$ symmetry of the GSWSP energy and examine the solution in the negative region. We define the following abbreviations
\begin{eqnarray}
-{\epsilon}^2 &\equiv& \frac{\big(E+ m_0c^2\big)\big(E- m_0c^2 \big)}{a^2\hbar^2c^2},\\
{\beta}^2  &\equiv& \frac{2\big(E+ m_0c^2\big)\big(V_0-W_0\big)}{a^2\hbar^2c^2},  \\
{\gamma}^2    &\equiv& \frac{2\big(E+m_0c^2\big)W_0}{a^2\hbar^2c^2}.
\end{eqnarray}
and we obtain
  \begin{eqnarray}
 \frac{d^2\phi_L(x)}{dx^2}+a^2\Bigg[-\epsilon^2+\frac{\beta^2}{1+e^{-a(x+L)}}+\frac{\gamma^2}
  {\big(1+e^{-a(x+L)}\big)^2}\Bigg] \phi_L(x) &=& 0. \label{KG2x<0}
\end{eqnarray}
We employ the continuity conditions on the wave functions followed by their asymptotic behaviour examinations.  Then, we find the following transcendental  equations
\begin{eqnarray}
\frac{(S_1 N_1)+(1-t_0)^{-2\nu}(S_2 N_2)}{(S_3 N_3)+(1-t_0)^{-1-2\nu}(S_4 N_4)}&=&  -\frac{(\mu+\theta+\nu)(1+\mu-\theta+\nu)t_0(t_0-1)}
{(1+2\mu)\big((\mu+\nu)t_0-\mu\big)}, \label{ciftcozum}\\
\frac{S_1 N_1}{S_2 N_2} &=& -(1-t_0)^{-2\nu}. \label{tekcozum}
\end{eqnarray}
for the calculation of the quantized energy spectrum with even and odd node numbers, respectively. Here, we use the definitions of $S_i$
\begin{eqnarray}
  S_1 &\equiv& \frac{\Gamma(1+2\mu)\Gamma(-2\nu)}{\Gamma(1+\mu-\theta-\nu)\Gamma(\mu+\theta-\nu)}, \\
  S_2 &\equiv& \frac{\Gamma(1+2\mu)\Gamma(2\nu)}{\Gamma(1+\mu-\theta+\nu) \Gamma(\mu+\theta+\nu)}, \\
  S_3 &\equiv& \frac{\Gamma(2+2\mu)\Gamma(-1-2\nu)}{\Gamma(1+\mu-\theta-\nu)\Gamma(\mu+\theta-\nu)}, \\
  S_4 &\equiv& \frac{\Gamma(2+2\mu)\Gamma(1+2\nu)}{\Gamma(2+\mu-\theta+\nu)\Gamma(1+\mu+\theta+\nu)},
\end{eqnarray}
and  $N_i$
\begin{eqnarray}
  N_1 &\equiv&  \,\,\, {}_2F_1[\mu+\theta+\nu,1+\mu-\theta+\nu,1+2\nu;1-t_0],  \\
  N_2 &\equiv&  \,\,\, {}_2F_1[1+\mu-\theta-\nu,\mu+\theta-\nu,1-2\nu;1-t_0],  \\
  N_3 &\equiv&  \,\,\, {}_2F_1[1+\mu+\theta+\nu,2+\mu-\theta+\nu,2+2\nu;1-t_0],  \\
  N_4 &\equiv&  \,\,\, {}_2F_1[1+\mu-\theta-\nu,\mu+\theta-\nu,-2\nu;1-t_0].
\end{eqnarray}
where $i=1,...,4$. Note that, we define $\mu$, $\nu$ and $\theta$ as follows:
\begin{eqnarray}
\mu&\equiv&\frac{1}{a \hbar c}\sqrt{-\big(E+ m_0c^2\big)\big(E- m_0c^2 \big)}\,,\label{k1} \\
\nu &\equiv& \frac{i}{a \hbar c} \sqrt{\big(E+ m_0c^2\big)\big(E- m_0c^2+2V_0 \big)}\,, \label{k2}\\
\theta &\equiv& \frac{1}{2}\mp \sqrt{\frac{1}{4}- \frac{2\big(E+m_0c^2\big)W_0}{a^2\hbar^2c^2}}\,,
\end{eqnarray}
and
\begin{eqnarray}
  t_0 \equiv \frac{1}{(1+e^{-a L})}.
\end{eqnarray}
As a final remark, we use the appropriate potential parameters that satisfy $e^{a L}>>1$ condition in this paper.

\section{Calculated Energy Spectra}\label{sec:EnergySpectra}
In this section, we assume a neutral pion is confined in the GSWSP energy wells that own either repulsive or attractive surface interactions. We use the Newton-Raphson numerical method to solve the transcendental equations given in Eq.~(\ref{ciftcozum}) and Eq.~(\ref{tekcozum}).  Note that, the rest mass energy of the neutral pion is $134.976$ $MeV$. In the case of repulsive surface effects, we assign a value of the twice of the mass energy of neutral pion to the parameter $W_0$. In the case that attractive surface interactions exist, we give the negative of the same value to the parameter $W_0$.

\subsection{Repulsive case}\label{subsec:Repulsive}
In Table \ref{tab1} we tabulate the calculated energy spectra  by changing the slope parameter while keeping the effective length parameter constant. We plot three GSWSP energy wells versus the distance in the fig. \ref{potfig03} (a). The increase in the slope parameter modifies the GSWSP energy wells in such a way that the distances between the center and the points where the value of the GSWSP energy well tends to zero, increase. These distances are calculated to be $4.9014$ $fm$, $5.2676$ $fm$, and $5.4507$ $fm$.  Furthermore, the wells reach their maximum height value of $37.962$ $MeV$,  at the points $6.5108$ $fm$, $6.3406$ $fm$ and $6.2554$ $fm$ far from the center.

\begin {table}[!ht]
\caption{\label{tab1}
Calculated energy spectra with the parameters $V_0=\frac{m_0c^2}{2}$, $W_0=2m_0c^2$ and $L=6$ $fm$. Note that the unit of energy eigenvalues is $MeV$.}
\begin{tabular}{|c|c|c|c|}
  \hline
                            &     $E_0$              & $E_1$              &  $E_2$                \\ \hline
  $a=1.0$ $fm^{-1}$         &     30.7239            & 82.4986            &  131.9274             \\ \hline
  $a=1.5$ $fm^{-1}$         &     21.1451            & 64.9900            &  112.7926             \\ \hline
  $a=2.0$ $fm^{-1}$         &     17.9094            & 57.5129            &  101.5504             \\ \hline
\end{tabular}
\end{table}

Then we increase the value of the parameter $V_0$ from $V_0=\frac{m_0c^2}{2}$ to $m_0c^2$. The increase of the volume effect parameter turns the GSWSP energy wells into deeper ones. The height of the barrier decreases to $16.872$ $MeV$. We plot three GSWSP energy wells with different slope parameters in  fig. \ref{potfig03} (c). Although the slope parameters increase, the distance of the point where the GSWSP energy is zero to the center remains the same. We calculate them as  $6.0000$ $fm$. Unlike these distances, other distances where the barriers reach their maximum value $16.872$ $MeV$ to the center change. They become $7.0986$ $fm$, $6.7324$ $fm$, and $6.5493$ $fm$, respectively. We tabulate the calculated eigenvalues in Table \ref{tab2}. As expected in deeper wells, the number of the eigenstates increase.

\begin {table}[!ht]
\caption{\label{tab2}
Calculated energy spectra with the parameters $V_0=m_0c^2$, $W_0=2m_0c^2$ and $L=6$ $fm$. Note that the unit of energy eigenvalues is $MeV$.}
\begin{tabular}{|c|c|c|c|c|c|}
  \hline
                        &   $E_0$      &   $E_1$       &   $E_2$      &  $E_3$     &  $E_4$      \\ \hline
  $a=1.0$ $fm^{-1}$     &  -75.9057    &  -10.7144     &   46.3186    &  97.0587   &   -         \\ \hline
  $a=1.5$ $fm^{-1}$     &  -85.6202    &  -27.9652     &   27.3165    &  78.9377   &  125.6045   \\ \hline
  $a=2.0$ $fm^{-1}$     &  -89.1376    &  -35.3352     &   18.0383    &  69.2923   &  117.0051   \\ \hline
\end{tabular}
\end{table}

Then, we decide to analyze the effect of the other potential parameter, $L$. We start with the GSWSP energy wells that are constructed with the depth parameter  value $\frac{m_0c^2}{2}$. We keep the slope parameter constant and increase the value of $L$. We plot the used wells in fig \ref{potfig04} (a). We notice that the increase of $L$, results with the wider GSWSP energy wells. Therefore, the number of the possible quantum states increase. The calculated spectra are given in Table \ref{tab3}. Another effect of the increase of the parameter $L$ is the increase of the distances where the potential energies reach the value of zero. They are calculated to be  $4.9014$ $fm$, $5.9014$ $fm$ and $5.9014$ $fm$. In addition, the distances where the particles feel the maximum barrier effects are found to be $6.5108$ $fm$,$7.5108$ $fm$ and $8.5108$ $fm$. Note that the calculated increases are same the proposed increased to the parameter.

\begin {table}[!ht]
\caption{\label{tab3}
Calculated energy spectra with the parameters $V_0=\frac{m_0c^2}{2}$, $W_0=2m_0c^2$ and $a=1$ $fm^{-1}$. Note that the unit of energy eigenvalues is $MeV$.}
\begin{tabular}{|c|c|c|c|c|}
  \hline
                        &    $E_0$      & $E_1$     & $E_2$         &   $E_3$    \\ \hline
  $L=6$ $fm$            &    30.7239    & 82.4986   & 131.9274      &   -        \\ \hline
  $L=7$ $fm$            &    22.0214    & 63.7775   & 106.9181      &   -        \\ \hline
  $L=8$ $fm$            &    16.3705    & 50.3163   &  87.6618      &  124.5361  \\ \hline
\end{tabular}
\end{table}

We investigate the role of the parameter $L$ in a deeper GSWSP energy well with repulsive surface effects by increasing the value of $V_0$ parameter to $m_0c^2$. We plot three GSWSP energy wells with three different values $L$ in fig \ref{potfig04} (c). We calculate the distances of the center with the points where the potential energies vanish as  follows $6.0000$ $fm$, $7.0000$ $fm$ and $8.0000$ $fm$. We calculate the distances of the points where the potential energy barrier have the maximum strength to the origin of the wells. We obtain an increase, too. Here are the calculated values  $7.0986$ $fm$, $8.0986$ $fm$ and $9.0986$ $fm$. The effects of the increase of the parameter $L$, the GSWSP energy wells enlarge. Therefore the number of the calculated eigenvalues increase. We tabulate the calculated energy spectra in Table \ref{tab4}.
\begin {table}[!ht]
\caption{\label{tab4}
Calculated energy spectra with the parameters $V_0=m_0c^2$, $W_0=2m_0c^2$ and $a=1$ $fm^{-1}$. Note that the unit of energy eigenvalues is $MeV$.}
\begin{tabular}{|c|c|c|c|c|c|c|}
  \hline
                &  $E_0$    &  $E_1$     &  $E_2$      &$E_3$    & $E_4$    & $E_5$     \\   \hline
  $L=6$ $fm$    & -75.9057  & -10.7144   &  46.3186    &97.0587  &  -       &  -        \\   \hline
  $L=7$ $fm$    & -86.4575  & -31.3941   &  18.9452    &65.1468  & 107.3222 &  -        \\   \hline
  $L=8$ $fm$    & -93.8577  & -46.6052   &  -2.0589    &39.6484  & 78.7117  & 115.0039  \\   \hline
\end{tabular}
\end{table}


\subsection{Attractive case}\label{subsec:Attractive}
In this subsection, we perform a similar calculation of the energy spectra within the attractive case. In Table \ref{tab5} we tabulate the calculated energy spectra  by increasing the slope parameter for a constant value of the  effective length parameter. We plot three GSWSP energy wells versus the distance in the fig. \ref{potfig03} (b). The increase in the slope parameter modifies the GSWSP energy wells in such a way that the wells reach their minimum pocket depth height value of $-105.4500$ $MeV$,  at the points $5.4892$ $fm$, $5.6595$ $fm$ and $5.7446$ $fm$ far from the center. Note that, the first physical quantum state does not appear in the zero node.

\begin {table}[!ht]
\caption{\label{tab5}
Calculated energy spectra with the parameters $V_0=\frac{m_0c^2}{2}$, $W_0=-2m_0c^2$ and $L=6$ $fm$. Note that the unit of energy eigenvalues is $MeV$.}
\begin{tabular}{|c|c|c|c|c|c|c|}
  \hline
                      & $E_0$     & $E_1$      & $E_2$     & $E_3$    &  $E_4$   &  $E_5$     \\ \hline
  $a=1.0$ $fm^{-1}$   &     -     & -          & 34.8076   & 68.0564  &  98.1005 & 123.2749   \\ \hline
  $a=1.5$ $fm^{-1}$   &     -     & 5.0473     & 41.8567   & 79.6105  & 113.3589 &   -        \\ \hline
  $a=2.0$ $fm^{-1}$   &     -     & 13.4515    & 46.5117   & 84.9667  & 121.1452 &   -        \\ \hline
\end{tabular}
\end{table}

When we increase the volume effects to $m_0c^2$, the relative depth of the pockets to the volume decrease while the maximum depth of the potential energy well increases to $-151.848$ $MeV$. As the result of this relative decrease, ground state wave function with zero node occurs. On the other hand, the increase of the maximum depth results with a high number of eigenvalues in the spectrum as given in Table \ref{tab6}. The increase of the slope parameter modifies the distance of the maximal depth point to the center slightly as shown in fig. \ref{potfig03} (d). We calculate them to be $4.9014$ $fm$, $5.2676$ $fm$ and $5.4507$ $fm$, respectively.

\begin {table}[!ht]
\caption{\label{tab6}
Calculated energy spectra with the parameters   $V_0=m_0c^2$, $W_0=-2m_0c^2$ and $L=6$ $fm$. Note that the unit of energy eigenvalues is $MeV$.}
\begin{tabular}{|c|c|c|c|c|c|c|c|c|}
  \hline
                      &  $E_0$     &  $E_1$    &  $E_2$    & $E_3$   &$E_4$   &  $E_5$   &  $E_6$     &  $E_7$     \\ \hline
   $a=1.0$ $fm^{-1}$  & -107.4134  & -72.0512  & -30.0914  & 10.4123 &47.5805 & 81.0955  & 110.2205   & 131.7276   \\ \hline
   $a=1.5$ $fm^{-1}$  & -103.6645  & -66.6949  & -25.2378  & 17.3292 &58.0249 & 95.1166  & 123.3235   &   -        \\ \hline
   $a=2.0$ $fm^{-1}$  & -101.7209  & -63.2435  & -21.1125  & 21.1125 &63.3810 & 102.7846 & 133.7675   &   -        \\ \hline
\end{tabular}
\end{table}

We examine the effects of the changes of the effective length parameter to the energy spectrum in the attractive case, too. In Table \ref{tab7} we assign $\frac{m_0c^2}{2}$ value to the potential depth parameter and calculate the spectrum for three different increasing value of $L$ shown in fig. \ref{potfig04} (b). The attractive case spectrum changes are parallel to the repulsive case changes, i.e. the number of the eigenvalues increase, the values of the first physical eigenvalue decrease, the potential energy wells enlarge with a linear increase. Note that the non existence of the first two nodes of the wave function does not change since the shape of the well changes horizontally. The distance of the maximum depth point to the center is $5.4892$ $fm$, $6.4892$ $fm$, and $7.4892$ $fm$.

\begin {table}[!ht]
\caption{\label{tab7}
Calculated energy spectra with the parameters  $V_0=\frac{m_0c^2}{2}$, $W_0=-2m_0c^2$ and $a=1$ $fm^{-1}$. Note that the unit of energy eigenvalues is $MeV$.}
\begin{tabular}{|c|c|c|c|c|c|c|c|}
  \hline
                &    $E_0$   & $E_1$ & $E_2$   & $E_3$   & $E_4$   &  $E_5$   & $E_6$     \\ \hline
   $L=6$ $fm$   &     -      & -     & 34.8076 & 68.0564 & 98.1005 & 123.2749 &  -        \\ \hline
   $L=7$ $fm$   &     -      & -     & 27.1043 & 57.1156 & 85.2305 & 110.6422 & 130.7738  \\ \hline
   $L=8$ $fm$   &     -      & -     & 21.3261 & 48.1762 & 74.1839 &  98.6009 & 120.3550  \\ \hline
\end{tabular}
\end{table}

In a deeper GSWSP energy well, as illustrated in fig. \ref{potfig04} (d), we examine the increasing effects of the parameter $L$.  We summarize the calculated energy spectra  in Table \ref{tab8}. The spectrum possesses the highest number of available states when the GSWSP energy well becomes most deeper and larger as one can suggest easily. Note that the calculated distances of the maximum deep point coordinate to the origin are $4.9014$ $fm$, $5.9014$ $fm$, and $6.9014$ $fm$ which are less than the case $V_0=\frac{m_0c^2}{2}$. At this point, the results  differ from the repulsive case.

\begin {table}[!ht]
\caption{\label{tab8}
Calculated energy spectra with the parameters  $V_0=m_0c^2$, $W_0=-2m_0c^2$ and $a=1$ $fm^{-1}$. Note that the unit of energy eigenvalues is $MeV$.}
\begin{tabular}{|c|c|c|c|}
  \hline
                    &    $L=6$ $fm$    & $L=7$ $fm$  & $L=8$ $fm$     \\ \hline
   $E_0$            &     -107.4134    & -109.9017   &  -111.9362     \\ \hline
   $E_1$            &     -72.0512     & -79.1807    &  -84.6845      \\ \hline
   $E_2$            &     -30.0914     & -42.6974    &  -52.6125      \\ \hline
   $E_3$            &     10.4123      & -6.0797     &  -19.5258      \\ \hline
   $E_4$            &    47.5805       & 28.4075     &  12.3167       \\ \hline
   $E_5$            &     81.0955      & 60.3835     &  42.3895       \\ \hline
   $E_6$            &    110.2205      & 86.3978     &  70.5581       \\ \hline
   $E_7$            &    131.7276      & 115.1339    &  96.4777       \\ \hline
   $E_8$            &    -             & 133.4480    &  119.1636      \\ \hline
   $E_9$            &    -             & -           &  134.5754      \\ \hline
\end{tabular}
\end{table}

As a final remark, we would like to point out the spectra that are calculated with the parameter $V_0=\frac{m_0c^2}{2}$ posses only positive energy eigenvalues. When the $V_0= m_0c^2$  parameter are used, the calculated eigenvalues in spectra own values among the KG interval. We would like to remark these obtained results are in a full agrement with the conclusion given in our previous paper \cite{Lutfuoglu_2018}.

\section{Thermodynamic Properties of a System}\label{sec:Thermo}
In an equilibrium system to investigate the thermodynamic properties, we devote the central role to the partition function \cite{Book_Santra}.  Although the partition function is a function of the absolute temperature, $T$, and microstates energies, $E_n$, in this work, we prefer to use its other definition given as
\begin{eqnarray}
  Z(\beta) &\equiv& \sum_{n} e^{-\beta E_n}.
\end{eqnarray}
Here, \emph{inverse temperature}, $\beta$, is defined by $\beta^{-1}\equiv k_B T$ and the Boltzmann constant, $k_B$, is equal to $8.6173 \times 10^{-11} MeV/K$.

Once the partition function is found, we can calculate the measure of the obtainable "useful" work in this closed thermodynamic system via the following relation
\begin{eqnarray}
  F(\beta) &=& - \frac{1}{\beta}\ln Z(\beta).
\end{eqnarray}
The well-known thermodynamic function entropy, $S(\beta)$, is derived from the Helmholtz free energy function as follows.
\begin{eqnarray}
  S(\beta) &=& - k_B \frac{\partial }{\partial \beta}F(\beta).
\end{eqnarray}
The internal energy of the system, $U(\beta)$, is the expectation energy value of the system. Mathematically, it is expressed with
\begin{eqnarray}
  U(\beta) &=& - \frac{\partial }{\partial \beta}\ln Z(\beta).
\end{eqnarray}
In an isochoric process, the specific heat function, $C_v(\beta)$, is the measure of the required energy to raise the absolute temperature $1$ $K$ per unit mass. It is derived by
\begin{eqnarray}
  C_v(\beta) &=& k_B \frac{\partial}{\partial \beta} U(\beta).
\end{eqnarray}

\section{Discussion}\label{sec:Discuss}
\begin{figure}[!htb]
\centering
\includegraphics[totalheight=0.45\textheight,clip=true]{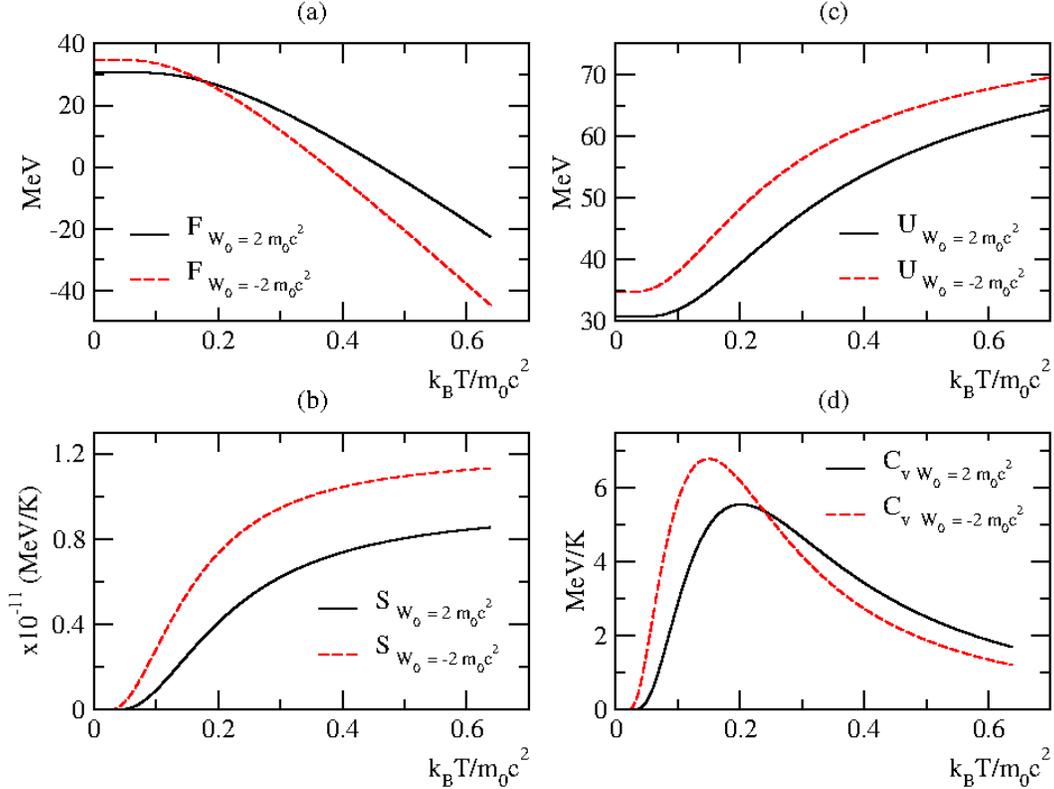}
   \caption{Thermodynamic functions of a confined neutral pion particle in either repulsive (solid-black lines) or attractive (dashed-red lines) GSWSP energy wells versus the reduced temperature. The other parameters of the wells are given as follows:  $V_0=\frac{m_0c^2}{2}$, $a=1$ $fm^{-1}$ and $L=6$ $fm$.} \label{fig1}
\end{figure}

In this section we obtain the thermodynamic functions versus the reduced temperature $\frac{k_BT}{m_0c^2}$ and then we compare them by classifying via the potential parameters. We start the discussion with a comparison via  $V_0$ parameter. Initially, we examine the confinement of the neutral pion in the wells  given in fig. \ref{potfig01} (a). The calculated energy spectra are given in the first row of the Table \ref{tab1} and Table \ref{tab5}. We demonstrate the thermodynamic functions in fig. \ref{fig1}.

The number of the calculated energy eigenvalues is higher in the attractive case. Therefore, the entropy function saturates at a higher value and the Helmholtz free energy decrease for high reduced temperature. The internal energies in the repulsive case start with the values of $30. 7239$ $MeV$ and saturates at $81.7166$ $MeV$, while in the attractive case launch with  $34.8076$ $MeV$ and saturate at $81.0601$ $MeV$.  We observe that in the attractive case, the specific heat reaches its maximum value at a lower reduced temperature.

\begin{figure}[!htb]
\centering
\includegraphics[totalheight=0.45\textheight,clip=true]{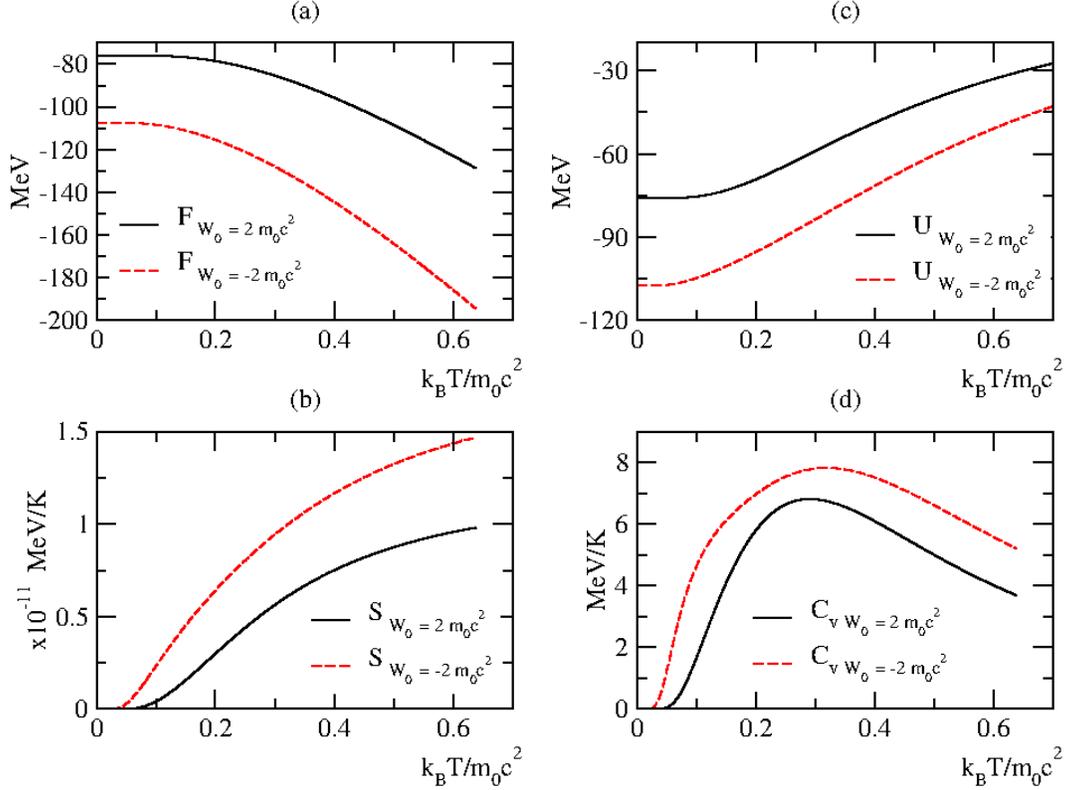}
   \caption{Thermodynamic functions of a confined neutral pion particle in either repulsive (solid-black lines) or attractive (dashed-red lines) GSWSP energy wells versus the reduced temperature. The other parameters of the wells are given as follows:  $V_0=m_0c^2$, $a=1$ $fm^{-1}$ and $L=6$ $fm$.} \label{fig2}
\end{figure}

Next, we compare the thermodynamic functions in the deeper GSWSP energy wells shown in fig. \ref{potfig01} (d). We use the calculated energy spectra given in the first rows of Tables \ref{tab2} and \ref{tab6}. The obtained thermodynamic functions are plotted in fig. \ref{fig2}.

The number of quantum microstates increased only in the repulsive case. The highest value of the Helmholtz free energies in the repulsive case is greater than the attractive case. From this point of view, the results differ from the shallow wells. We observe that the saturation value of the entropy function is higher in the deeper well. The average of the internal energies is calculated to be $14.1893$ $MeV$ and $21.4351$ $MeV$  in the repulsive and attractive cases, respectively. Finally, the highest value of the specific heat is reached in the greater value of the reduced temperature in the attractive case compared with the repulsive case.

\begin{figure}[!htb]
\centering
\includegraphics[totalheight=0.45\textheight,clip=true]{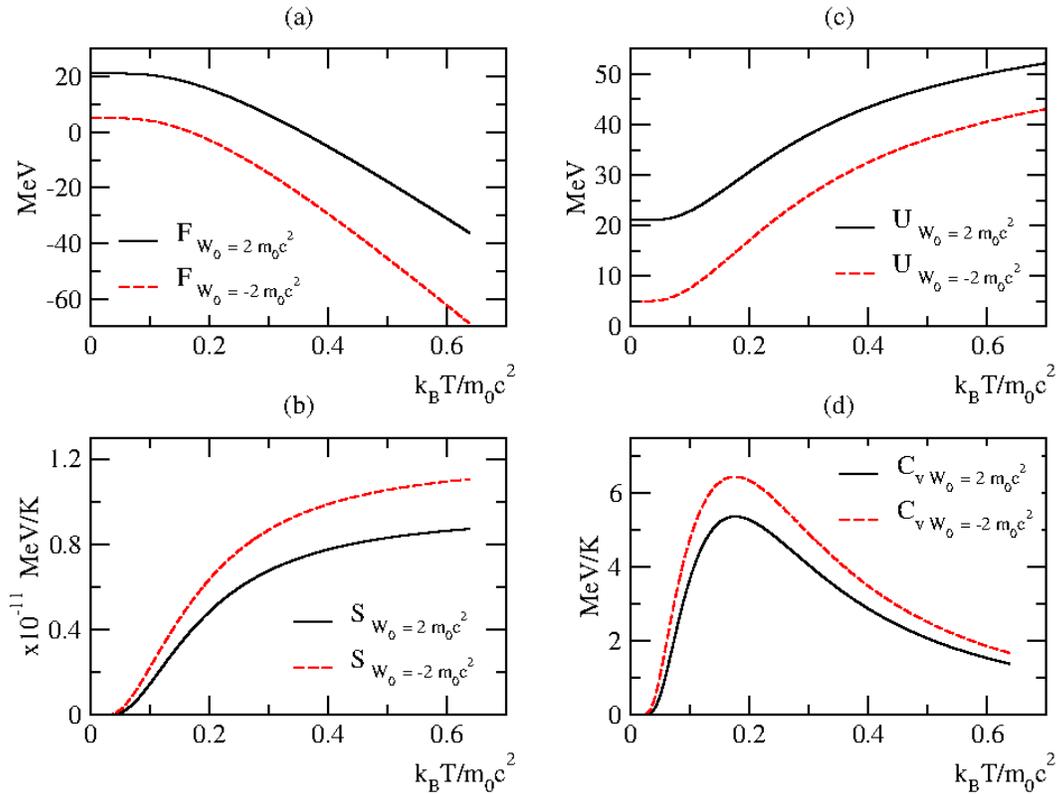}
   \caption{Thermodynamic functions of a confined neutral pion particle in either repulsive (solid-black lines) or attractive (dashed-red lines) GSWSP energy wells versus the reduced temperature. The other parameters of the wells are given as follows:  $V_0=\frac{m_0c^2}{2}$, $a=1.5$ $fm^{-1}$ and $L=6$ $fm$.} \label{fig3}
\end{figure}

\newpage
Next, we decide to investigate the effects of the changes in the slope parameter on the thermodynamic functions. Initially, we will investigate in the shallow wells, then in the deeper wells. Although the increase of the slope parameter from $1$ $fm^{-1}$ to $1.5$ $fm^{-1}$ has resulted in the widening of the wells, the number of the energy eigenvalues does not change.  We demonstrate the investigated wells in fig. \ref{potfig01} (b). We obtain the thermodynamic functions through the use of the calculated energy spectra given in the second rows of Tables \ref{tab1} and \ref{tab5}. We plot the  thermodynamic functions in fig. \ref{fig3}.

We observe that the highest value of the Helmholtz free energy of the repulsive case is higher than the attractive case. The relative decrease of the value of the spectra energies results with a slight decrease of the saturation value of the entropies. We calculate the saturation values of the mean energies decrease to $66.9900$ $MeV$ in the repulsive case and $59.9684$ $MeV$ in the attractive case. In both case, the highest value of the specific heat are reached nearly in the same value of the reduced temperature.

\begin{figure}[!htb]
\centering
\includegraphics[totalheight=0.45\textheight,clip=true]{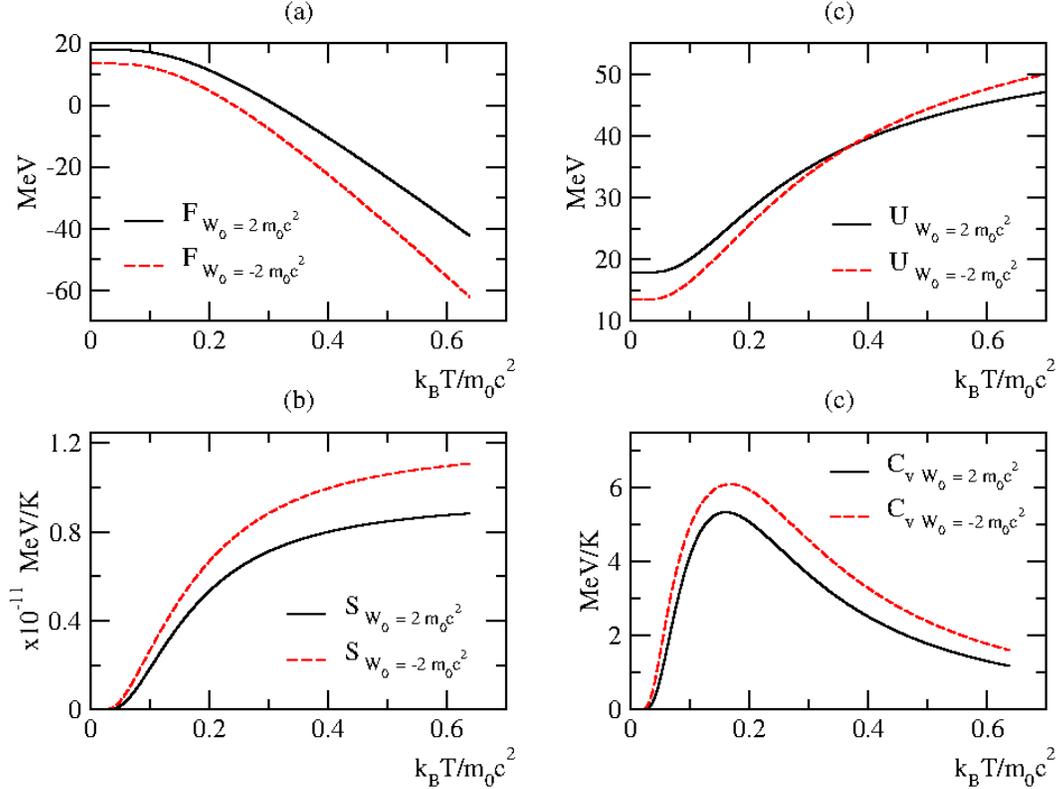}
   \caption{Thermodynamic functions of a confined neutral pion particle in either repulsive (solid-black lines) or attractive (dashed-red lines) GSWSP energy wells versus the reduced temperature. The other parameters of the wells are given as follows:   $V_0=\frac{m_0c^2}{2}$, $a=2$ $fm^{-1}$ and $L=6$ $fm$.} \label{fig4}
\end{figure}

\newpage
Then, we use a higher value of the slope parameter, i.e. $2$ $fm^{-1}$,  in the shallow wells. The ground level of the potential wells becomes wider since the slope increased as shown in fig. \ref{potfig01} (c). The wells with attractive and repulsive surface effects become more similar relatively, thus, the values of the calculated energy eigenvalues. These values are given in the third rows of Tables \ref{tab1} and \ref{tab5}. We illustrate the thermodynamic functions in fig. \ref{fig4}.

We observe that the difference of the Helmholtz free energies decrease. Consequently, the entropy functions behave similarly. The main change is seen in the average of the internal energies. In the attractive case, it is calculated to be $66.5188$ $MeV$. In the repulsive case it is less, $58.9909$ $MeV$. We observe that the specific heats reach thier maximal values in a smaller value of the reduced temperature in the repulsive case.

\begin{figure}[!htb]
\centering
\includegraphics[totalheight=0.45\textheight,clip=true]{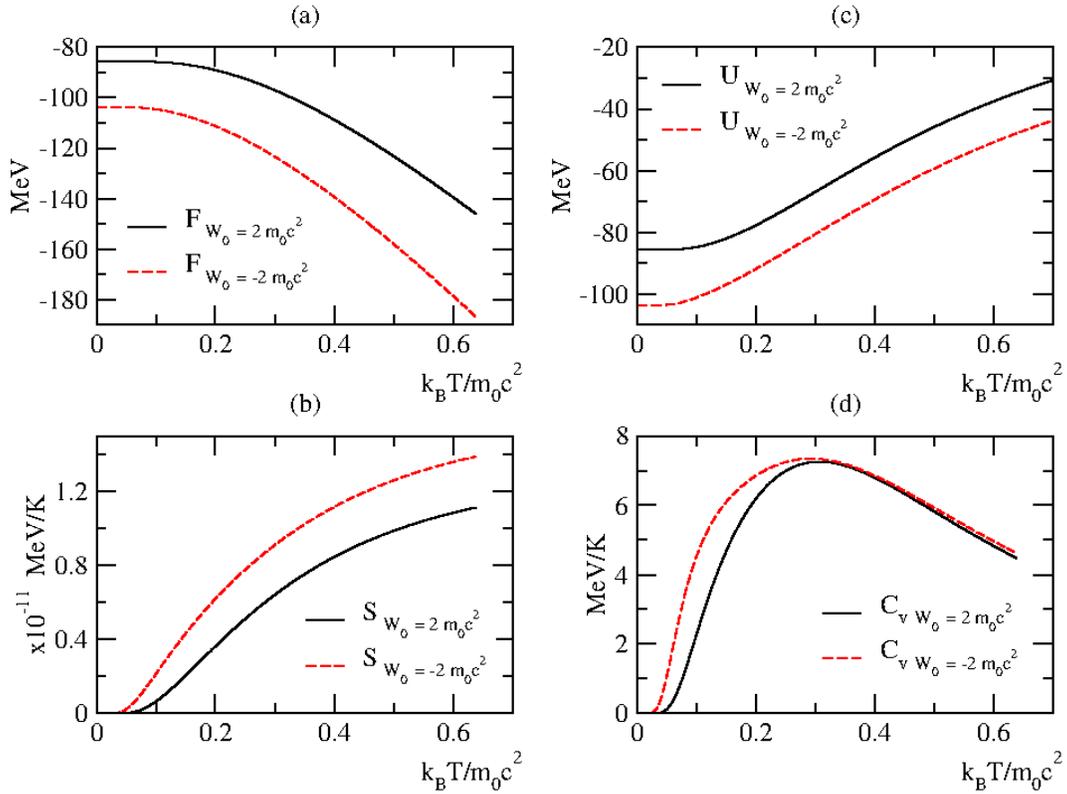}
   \caption{Thermodynamic functions of a confined neutral pion particle in either repulsive (solid-black lines) or attractive (dashed-red lines) GSWSP energy wells versus the reduced temperature. The other parameters of the wells are given as follows:   $V_0=m_0c^2$, $a=1.5$ $fm^{-1}$ and $L=6$ $fm$.} \label{fig7}
\end{figure}

\newpage
Next, we examine the same thermodynamic functions for the changes of the slope parameter in the deeper GSWSP energy wells. The increase of the slope parameter has the similar effect in means of the increase of the length of the ground of the wells. Note that, in a deeper well the surface effects weaken. Therefore, the potential energies become more and more similar. We start with $a=1.5$ $fm^{-1}$ and plot the potential wells in fig. \ref{potfig01} (e). We use the calculated energy spectra given in the second rows of Tables \ref{tab2} and \ref{tab6}. We note that the number of eigenvalues increase in the repulsive case and decrease in the attractive case. This is one of the major difference we observe. We demonstrate the thermodynamic functions in fig. \ref{fig7}.

The Helmholtz free energy starts from a higher energy value in the repulsive case. The difference between the Helmholtz free energy grow while the reduce temperature increase. The entropy functions saturate at a lower value compared with the case discussed in fig. \ref{fig6}. The average values of the internal energies increased to $23.6547$ $MeV$ in the repulsive case, while decreased to $14.0281$ $MeV$.

\begin{figure}[!htb]
\centering
\includegraphics[totalheight=0.45\textheight,clip=true]{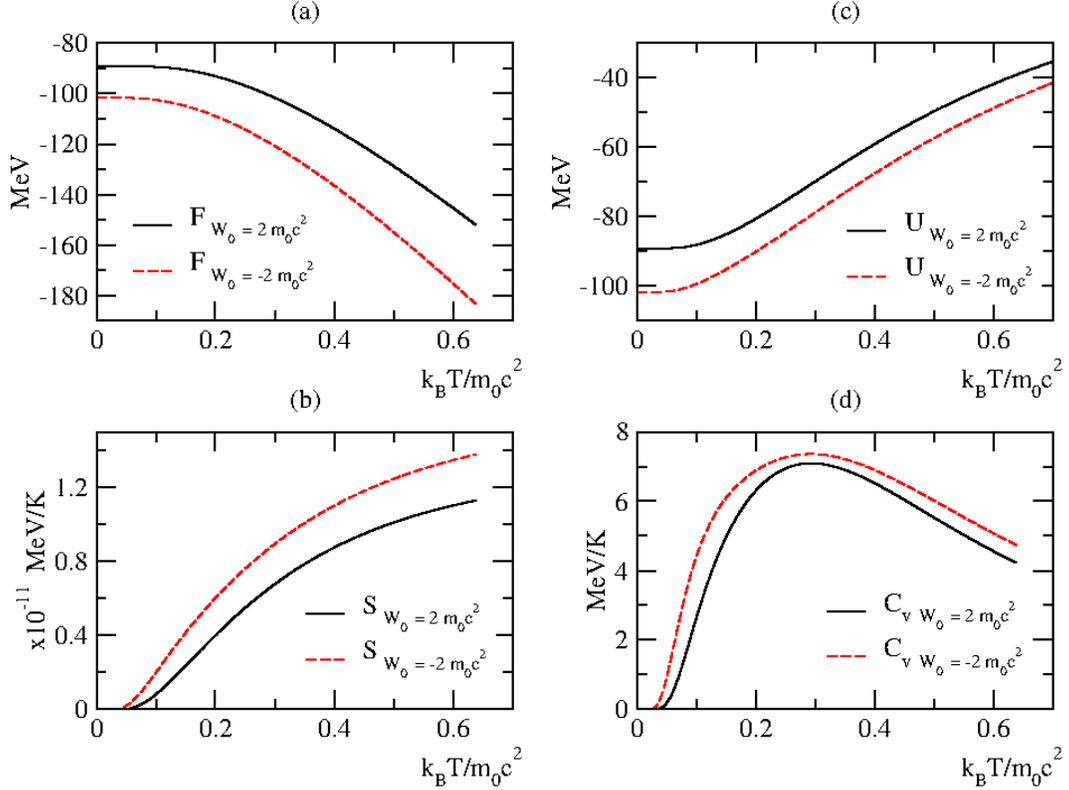}
   \caption{Thermodynamic functions of a confined neutral pion particle in either repulsive (solid-black lines) or attractive (dashed-red lines) GSWSP energy wells versus the reduced temperature. The other parameters of the wells are given as follows:   $V_0=m_0c^2$, $a=2$ $fm^{-1}$ and $L=6$ $fm$.} \label{fig8}
\end{figure}

Then, we investigate the thermodynamic functions for the highest assigned value of the slope parameter. The plot of the potential energy wells are given in fig. \ref{potfig01} (f). In this case, we obtain "most similar" potential energy wells in this paper. Consequently, we find the spectra are very "similar" to each other. We tabulate the calculated energy spectra in the third rows of Tables \ref{tab2} and \ref{tab6}. We plot four thermodynamic functions in fig. \ref{fig8}.

All thermodynamic functions that we examine, become closer. The internal energies saturates in a less value, $15.9726$ $MeV$, in the repulsive case. Moreover, in the attractive case the average value of the internal energy increases to $19.1790$ $MeV$.

We conclude that when the volume effects are more dominant, the increase of the slope parameter results the thermodynamic functions' collapse.

\begin{figure}[!htb]
\centering
\includegraphics[totalheight=0.45\textheight,clip=true]{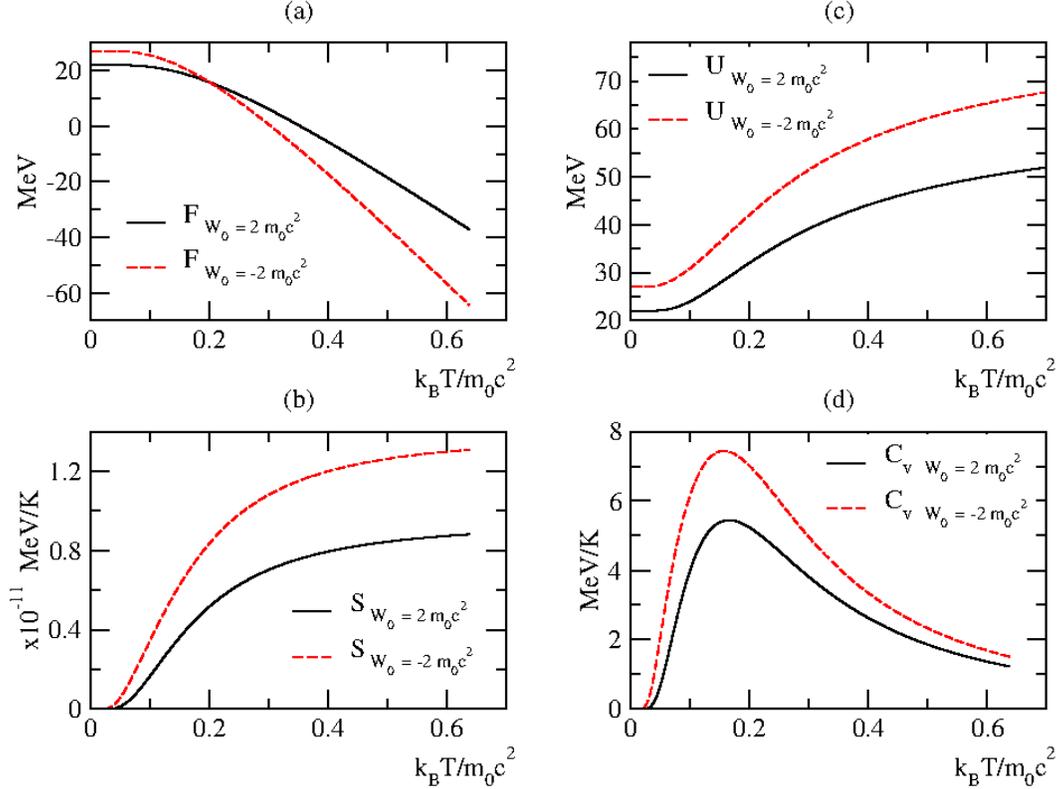}
   \caption{Thermodynamic functions of a confined neutral pion particle in either repulsive (solid-black lines) or attractive (dashed-red lines) GSWSP energy wells versus the reduced temperature. The other parameters of the wells are given as follows:   $V_0=\frac{m_0c^2}{2}$, $a=1$ $fm^{-1}$ and $L=7$ $fm$.} \label{fig5}
\end{figure}

\begin{figure}[!htb]
\centering
\includegraphics[totalheight=0.45\textheight,clip=true]{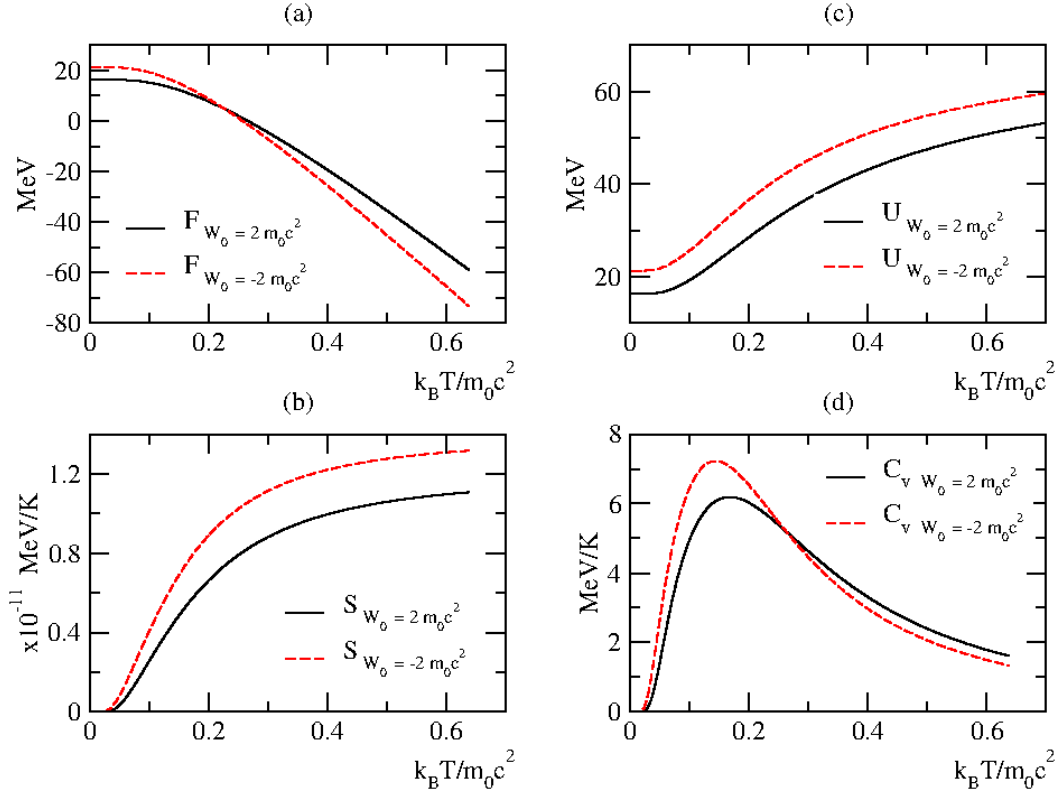}
   \caption{Thermodynamic functions of a confined neutral pion particle in either repulsive (solid-black lines) or attractive (dashed-red lines) GSWSP energy wells versus the reduced temperature. The other parameters of the wells are given as follows:   $V_0=\frac{m_0c^2}{2}$, $a=1$ $fm^{-1}$ and $L=8$ $fm$.} \label{fig6}
\end{figure}

\newpage
Another parameter that determines the shape of the GSWSP energy well is the effective length. Therefore, we examine the thermodynamic functions versus the changes of the parameter $L$. Initially, we start the investigation in the shallow wells. We demonstrate the GSWSP energy wells in fig. \ref{potfig02} (b). The effects of the increase extends the wells in both sides symmetrically. The calculated spectra are given in the second rows of Table \ref{tab3} and Table \ref{tab7}. The number of the eigenvalue does not increase in the repulsive case, but  in the attractive case. We present the plots of the thermodynamic functions in fig. \ref{fig5}.

Although the Helmholtz free energy of the attractive case initially starts from a higher value than the repulsive case, it decreases rapidly. The entropy functions behave similar, they are saturated in a lower value when they are compared with the effective length is $6$ $fm$. The average of the internal energies are found to be $64.2390$ $MeV$ and $82.1733$ $MeV$ in the repulsive and attractive cases, respectively. Specific heat functions has a higher extremum value in a lower reduced temperature in the attractive case.

Then, we investigate the same thermodynamic functions for a higher value of the effective length parameter.  We take $L=8$ $fm$ and illustrate the GSWSP energy wells in fig. \ref{potfig02} (c). The number of eigenvalues in the repulsive case increased due to the extension of the well. The spectra are given in the third rows of Tables \ref{tab3} and \ref{tab7}. Moreover, we demonstrate the obtained thermodynamic functions in fig. \ref{fig6}.

We observe that the characteristic of the thermodynamic functions does not change. The Helmholtz free energy function in the attractive case initially has a higher value than the one in the repulsive case. The increase in the number of the available quantum states increases the decreasing speed of the Helmholtz free energy when it is compared with the $L=7$ $fm$ case. We observe the resulting effect on the entropy function in the repulsive case such that the gap between two entropy functions is narrowed. We calculate the average internal energy as  $69.7212$ $MeV$ and $72.5284$ $MeV$ in the repulsive and attractive cases, respectively.

\begin{figure}[!htb]
\centering
\includegraphics[totalheight=0.45\textheight,clip=true]{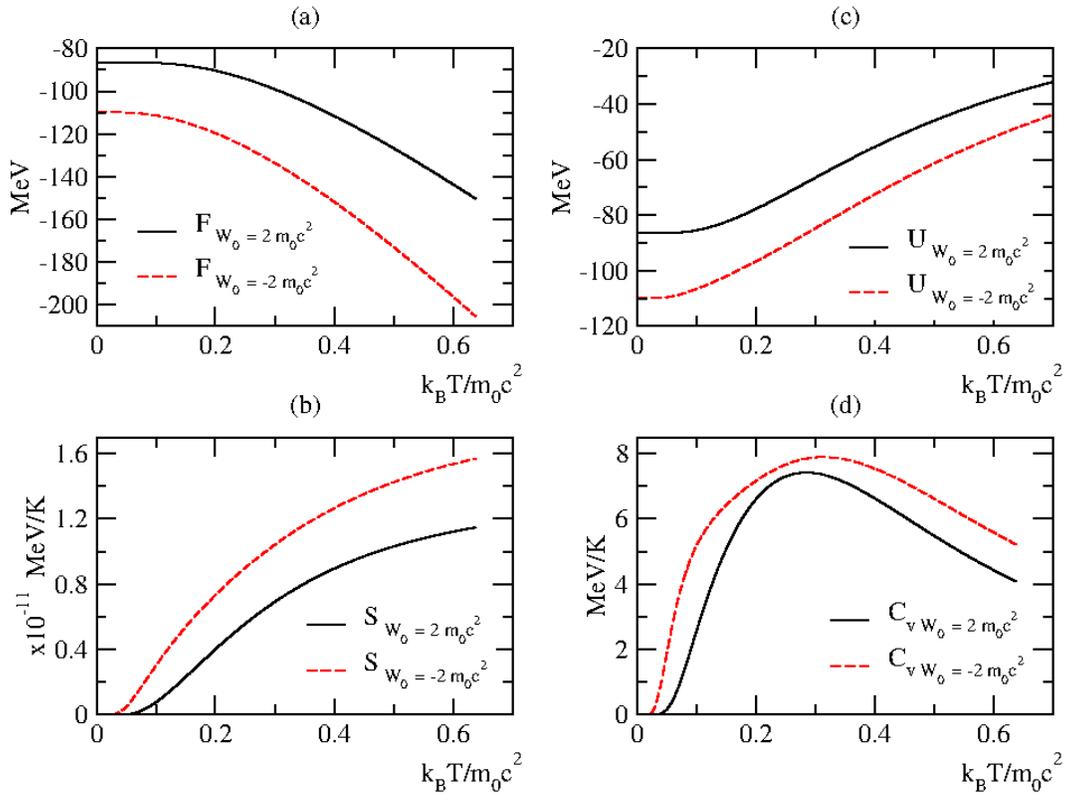}
   \caption{Thermodynamic functions of a confined neutral pion particle in either repulsive (solid-black lines) or attractive (dashed-red lines) GSWSP energy wells versus
    the reduced temperature. The other parameters of the wells are given as follows:   $V_0=m_0c^2$, $a=1$ $fm^{-1}$ and $L=7$ $fm$.} \label{fig9}
\end{figure}

\newpage
Next, we examine the same thermodynamic functions versus the changes of the effective length parameter in the deeper GSWSP energy wells. The increase of the parameter $L$ has the similar effect on the shape of the potential  as in the shallow wells, the wells extend without a change characteristically.  Initially, we start with $L=7$ $fm$. The plot of the potential wells is seen in fig. \ref{potfig02} (e). We use the calculated energy spectra given in the second row of Table \ref{tab4} and the second column of Table \ref{tab8}. In both case, the number of available quantum states increased by one in the energy spectra.  We demonstrate the thermodynamic functions in fig. \ref{fig9}.

The Helmholtz free energy function of the repulsive case starts from a higher value than the attractive case. When we compare the decreases of the functions, we observe a slower lessen in the repulsive case. The saturation values of the internal energies are $14.7125$ $MeV$ in the repulsive case and $20.6568$ $MeV$ in the attractive case.

\begin{figure}[!htb]
\centering
\includegraphics[totalheight=0.45\textheight,clip=true]{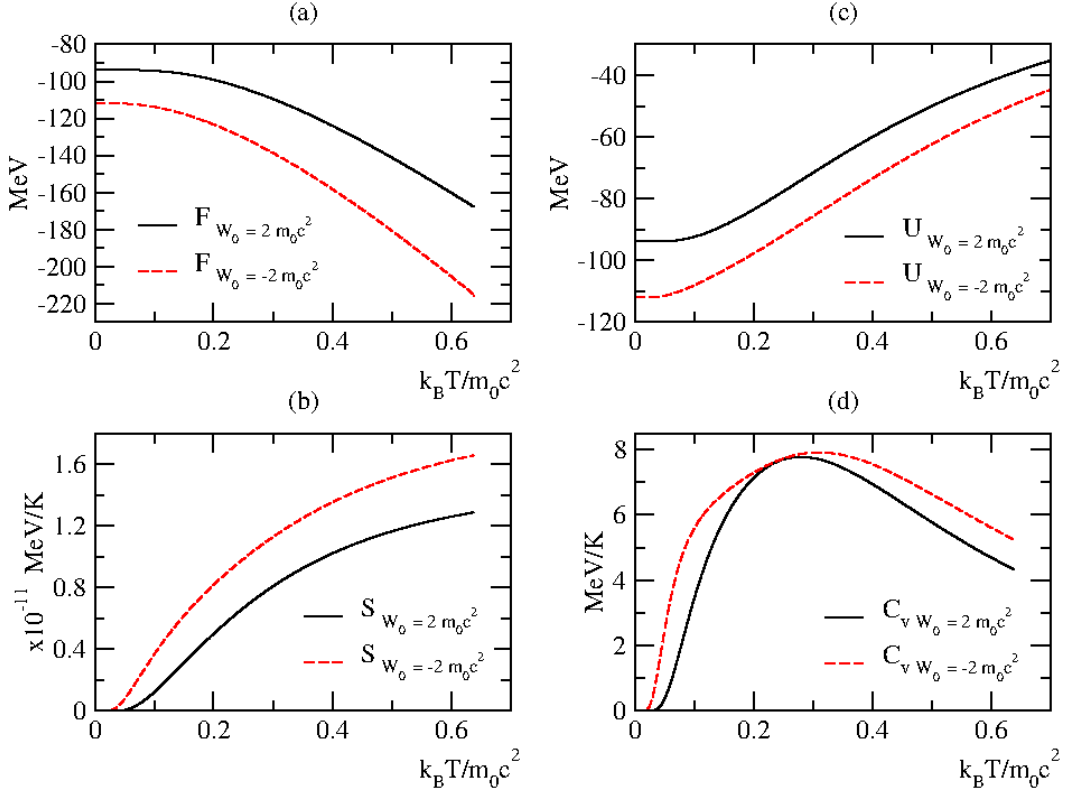}
   \caption{Thermodynamic functions of a confined neutral pion particle in either repulsive (solid-black lines) or attractive (dashed-red lines) GSWSP energy wells versus
    the reduced temperature. The other parameters of the wells are given as follows:   $V_0=m_0c^2$, $a=1$ $fm^{-1}$ and $L=8$ $fm$.} \label{fig10}
\end{figure}

\newpage
Then, we examine the thermodynamic functions in the potential energy wells that have longer effective length. We assign $8$ $fm$ value to the parameter $L$ and plot the GSWSP energy wells in \ref{potfig02} (f). Note that the calculated energy spectra are given in the third row of Table \ref{tab4} and the third column of Table \ref{tab8}. Alike the latter one,  the number of available quantum states increased by one in the energy spectra.  We illustrate the thermodynamic functions in fig. \ref{fig10}.

In both cases, we observe a faster decrease in the Helmholtz free energy and a faster increase in entropy functions when it is a comparison with the latter one. The average values of the internal energies are calculated to be $15.1404$ $MeV$ in the repulsive and $20.6722$ $MeV$ in the attractive cases, respectively. Note that the specific heat functions have same values at some critical reduced temperature values.

We conclude that the increase of the effective length does not create a characteristic change in the thermodynamic functions.

\begin{figure}[!htb]
\centering
\includegraphics[totalheight=0.45\textheight,clip=true]{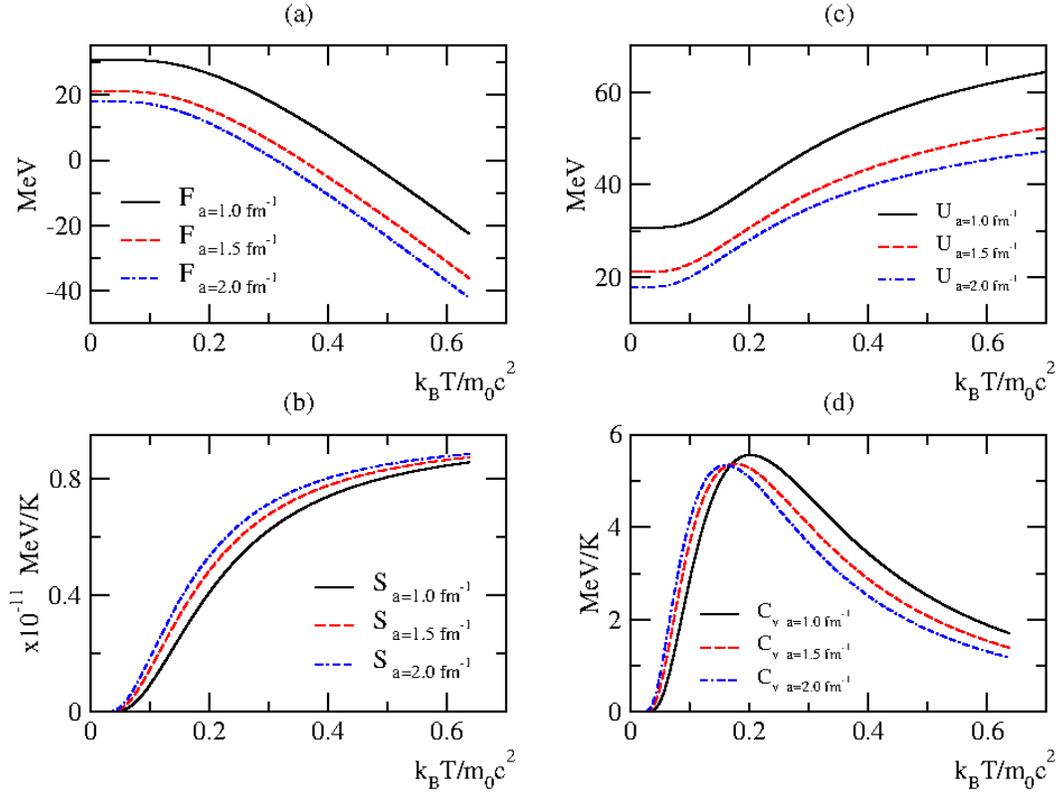}
   \caption{Reduced temperature versus the thermodynamic functions of a confined neutral pion particle in the repulsive GSWSP energy wells that are constructed with the parameters given as follows: $V_0=\frac{m_0c^2}{2}$, $W_0=2m_0c^2$ and $L=6$ $fm$.} \label{fig11}
\end{figure}

So far we have discussed the thermodynamic functions in the comparison between  the repulsive and attractive cases. In the remainder of this chapter, we want to examine the effect of shape parameters of potential energy on thermodynamic functions in the presence of repulsive or attractive surface effects. We start with the shallow well with repulsive surface effects. We plot three GSWSP energy wells with different values of the slope parameter in fig. \ref{potfig03} (a). We use the calculated spectra given in Table \ref{tab1}. We demonstrate the thermodynamic functions in fig. \ref{fig11}.

The Helmholtz free energy function decreases as a result of the increment of the slope parameter. Although the entropy function shows a faster increase, it becomes a saturation around a certain value. The function of the internal energy decreases before the characteristic is disturbed. Specific heat reaches a lower maximum at a lower reduced temperature. Note that these changes are not linear.

\begin{figure}[!htb]
\centering
\includegraphics[totalheight=0.45\textheight,clip=true]{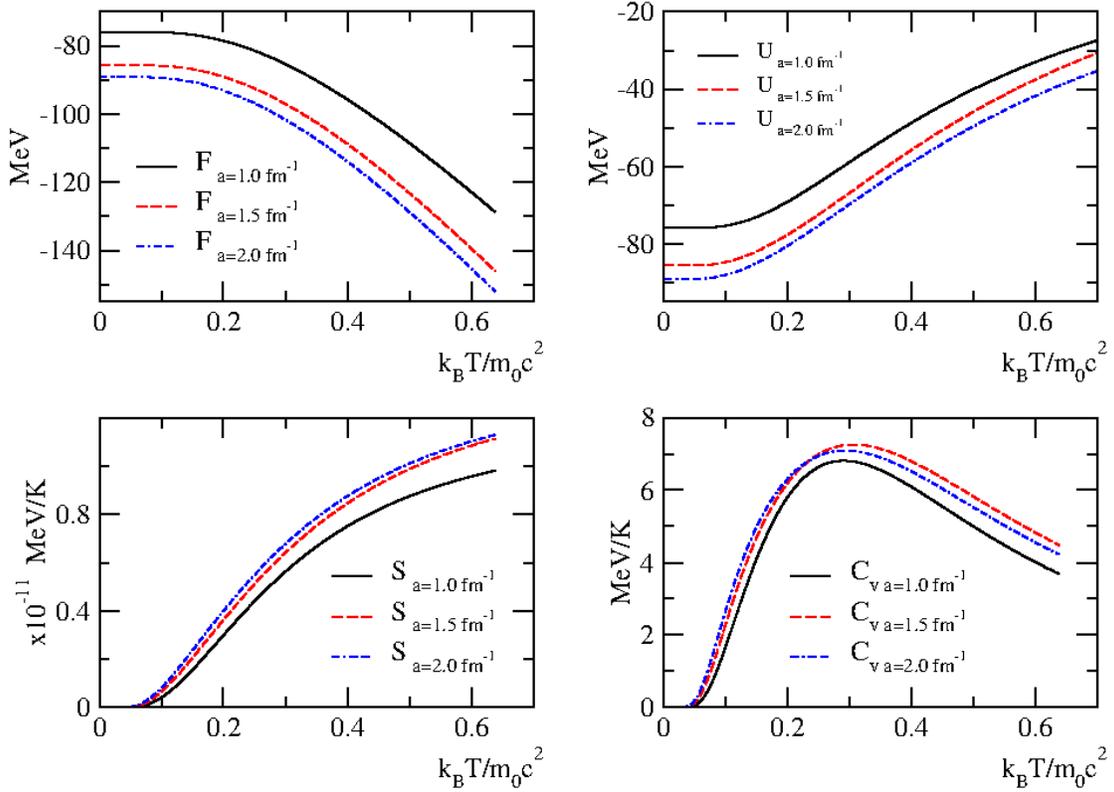}
   \caption{Reduced temperature versus the thermodynamic functions of a confined neutral pion particle in the repulsive GSWSP energy wells that are constructed with the parameters given as follows:   $V_0=m_0c^2$, $W_0=2m_0c^2$ and $L=6$ $fm$.} \label{fig12}
\end{figure}

\newpage
Then, we repeat the procedure for the deeper case. We plot three GSWSP energy wells with different values of the slope parameter in fig. \ref{potfig03} (c). We use the calculated spectra given in Table \ref{tab2}. We demonstrate the Helmholtz Free energy, entropy, internal energy and specific heat functions in fig. \ref{fig12}.

All thermodynamic functions except specific heat exhibit similar behavior as in the shallow well. The entropy function  almost overlaps when the slope parameter is $1.5$ $fm^{-1}$ and $2$ $fm^{-1}$, respectively. As the slope parameter increases in a deeper well, the peak value of the specific heat function increases. It is seen that these peak values are reached at greater reduced temperature values.

\begin{figure}[!htb]
\centering
\includegraphics[totalheight=0.45\textheight,clip=true]{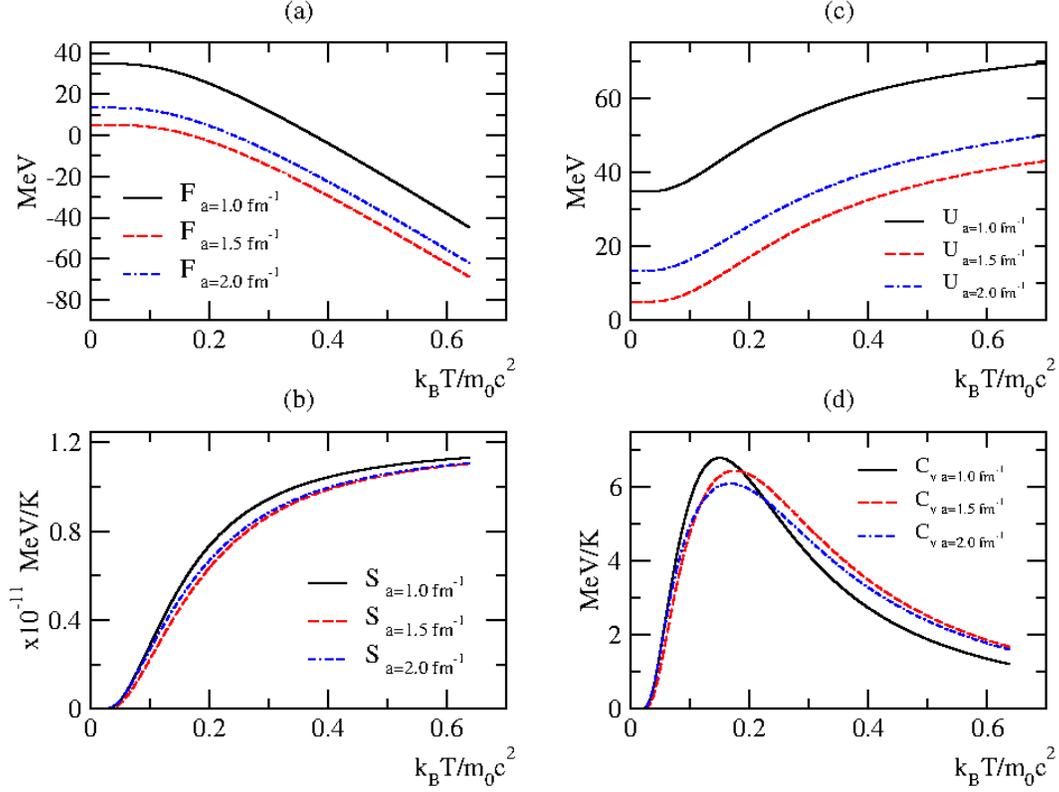}
   \caption{Reduced temperature versus the thermodynamic functions of a confined neutral pion particle in the attractive GSWSP energy wells that are constructed with the parameters given as follows:   $V_0=\frac{m_0c^2}{2}$, $W_0=-2m_0c^2$ and $L=6$ $fm$.} \label{fig13}
\end{figure}

\newpage
Next, we would like to examine the effect of the increase of slope parameter on the strong attractive surface effect in a shallow potential energy well. We demonstrate the GSWSP energy wells in fig. \ref{potfig03} (b). Calculated energy eigenvalues are given in Table \ref{tab5}. The obtained thermodynamic functions are illustrated in fig. \ref{fig13}.

The increase in the slope parameter appears to cause a decrease in the Helmholtz free energy entropy and internal energy functions and then an increase. This change causes the peak value of the specific heat function to be seen first at a greater, then at a slightly lower, reduced temperature. However, it is seen that the specific heat function's peak value decreases. We would like to emphasize that such a fluctuation is not seen in the effects of repulsive surface effects even in shallow or deep GSWSP energy wells.

\begin{figure}[!htb]
\centering
\includegraphics[totalheight=0.45\textheight,clip=true]{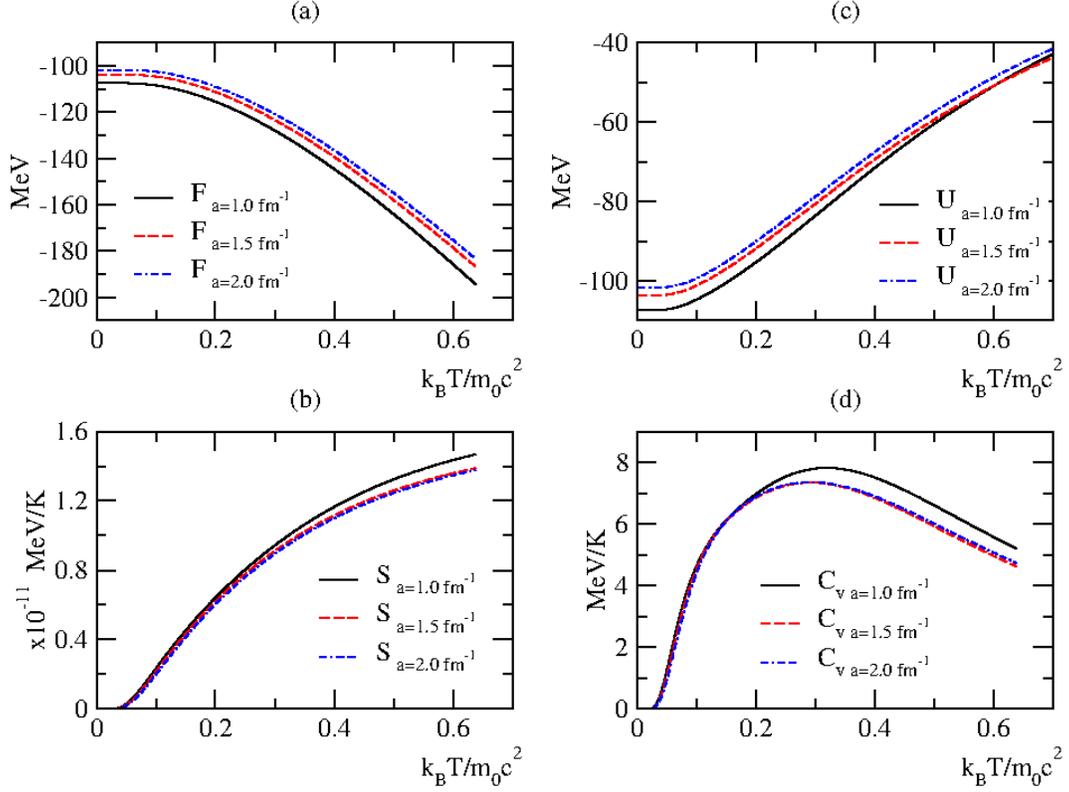}
   \caption{Reduced temperature versus the thermodynamic functions of a confined neutral pion particle in the attractive GSWSP energy wells that are constructed with the parameters given as follows:    $V_0=m_0c^2$, $W_0=-2m_0c^2$ and $L=6$ $fm$. } \label{fig14}
\end{figure}

\newpage
Then, we would like to examine the effect of the increase of slope parameter on strong surface effect in a deep potential energy well. We plot the GSWSP energy wells in fig. \ref{potfig03} (d). Calculated energy eigenvalues are tabulated in Table \ref{tab6}. The obtained thermodynamic functions are demonstrated in fig. \ref{fig14}.

The increase in the slope parameter appears to cause an increase in the Helmholtz free energy entropy and internal energy functions while a decrease in the entropy function and specific heat function. Moreover, the specific heat function's peak value decreases. Despite the increase in the slope parameter in a deep well, it overlaps the entropy and specific heat functions. This indicates the existence of a critical value.

\begin{figure}[!htb]
\centering
\includegraphics[totalheight=0.45\textheight,clip=true]{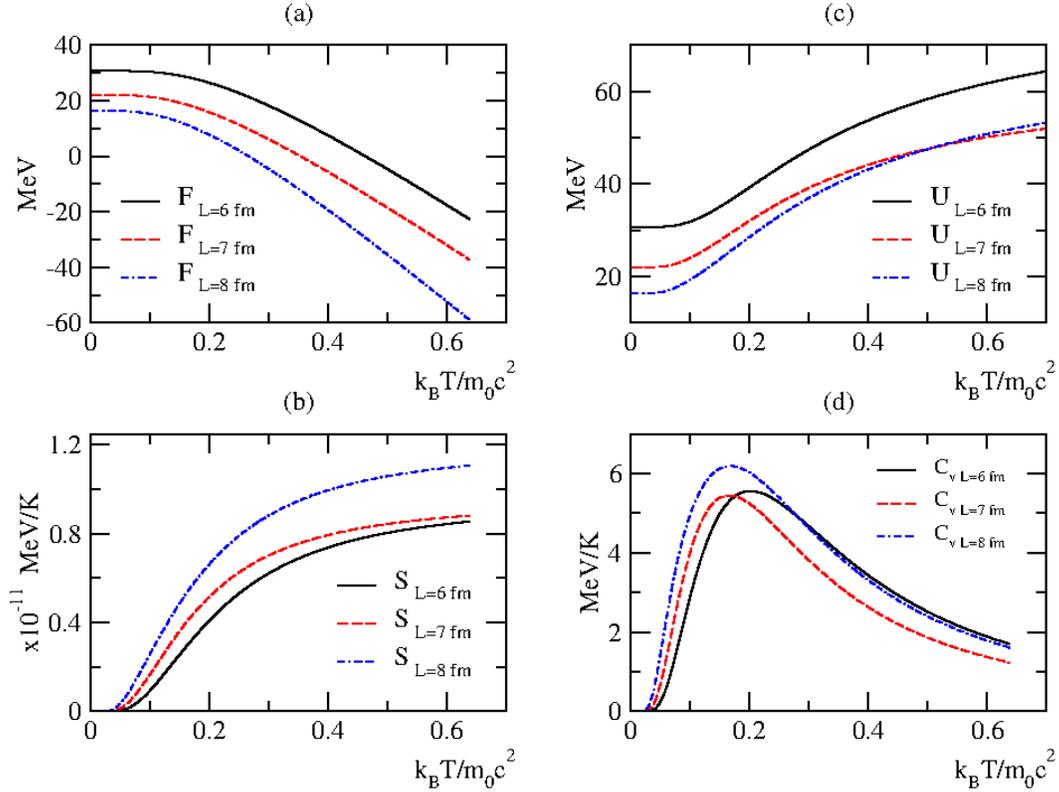}
   \caption{Reduced temperature versus the thermodynamic functions of a confined neutral pion particle in the repulsive GSWSP energy wells that are constructed with the parameters given as follows: $V_0=\frac{m_0c^2}{2}$, $W_0=2m_0c^2$ and $a=1$ $fm^{-1}$.} \label{fig15}
\end{figure}

\newpage
Finally, we would like to conclude by discussing the impact of the increase in effective length on the thermodynamic functions in the presence of repulsive and attractive surface effects in the shallow and the deeper wells. We start with the shallow well with repulsive surface effects. We plot three GSWSP energy wells with different values of the effective length parameter in fig. \ref{potfig04} (a). We use the calculated spectra given in Table \ref{tab3}. We demonstrate the thermodynamic functions in fig. \ref{fig15}.

When the effective distance receives the largest value we assign, the number of energy eigenvalues in the spectrum has increased. This increase in the number of enterable states provides a more rapid reduction in the Helmholtz free energy function. Therefore, entropy function reaches its saturation in a different value. The specific heat function exhibits almost the same behavior after a certain reduced temperature value of $ L = 6 $ $ fm $ and $ L = 8 $ $fm$.

\begin{figure}[!htb]
\centering
\includegraphics[totalheight=0.45\textheight,clip=true]{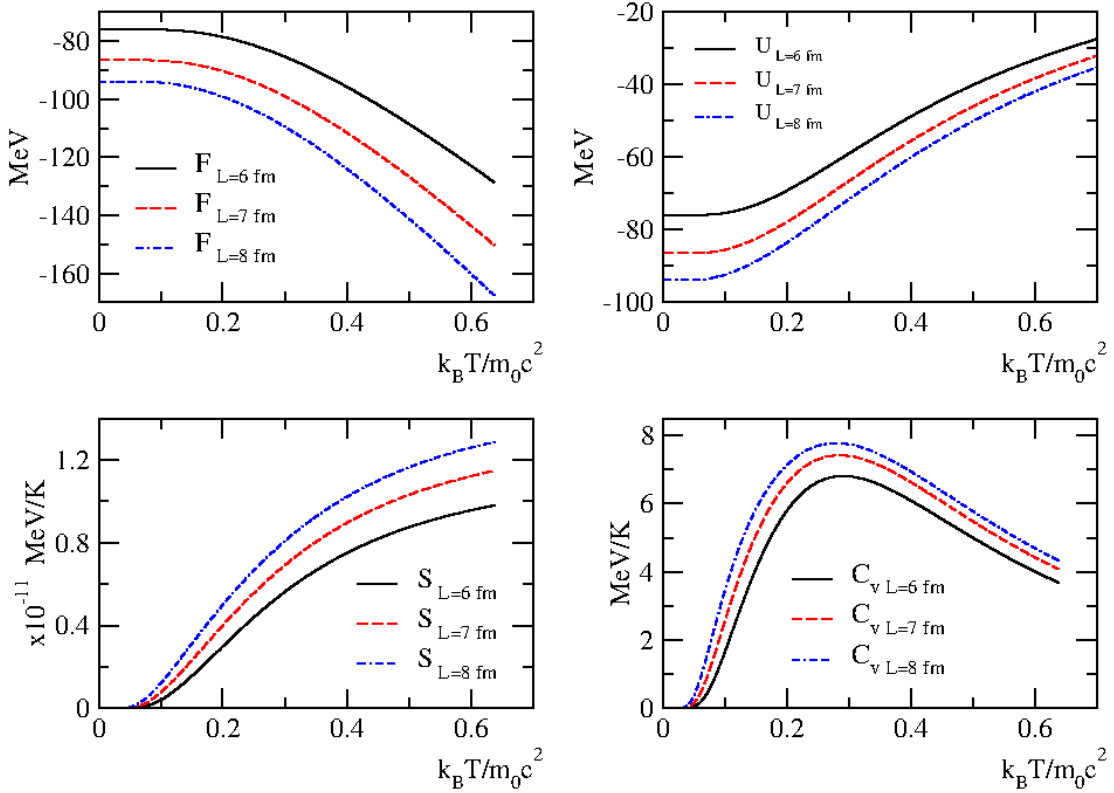}
   \caption{Reduced temperature versus the thermodynamic functions of a confined neutral pion particle in the repulsive GSWSP energy wells that are constructed with the parameters given as follows: $V_0=m_0c^2$, $W_0=2m_0c^2$ and $a=1$ $fm^{-1}$.} \label{fig16}
\end{figure}

\newpage
Then, we continue to investigate the impact of the increase in effective length on the thermodynamic functions in the presence of repulsive effects in the deeper wells.  We demonstrate three GSWSP energy wells via the increasing values of the effective length parameter in fig. \ref{potfig04} (c). Here, the calculated spectra tabulated in Table \ref{tab4} are used. The thermodynamic functions are illustrated in fig. \ref{fig16}.

Increasing the effective distance increases the number of quantum states that can be entered. Thus, Helmholtz free energy function starts to decrease rapidly. Entropy functions vary regularly. Although the internal energy function starts at different values, it reaches saturation at almost the same point. The specific heat function changes properly just like entropy.

\begin{figure}[!htb]
\centering
\includegraphics[totalheight=0.45\textheight,clip=true]{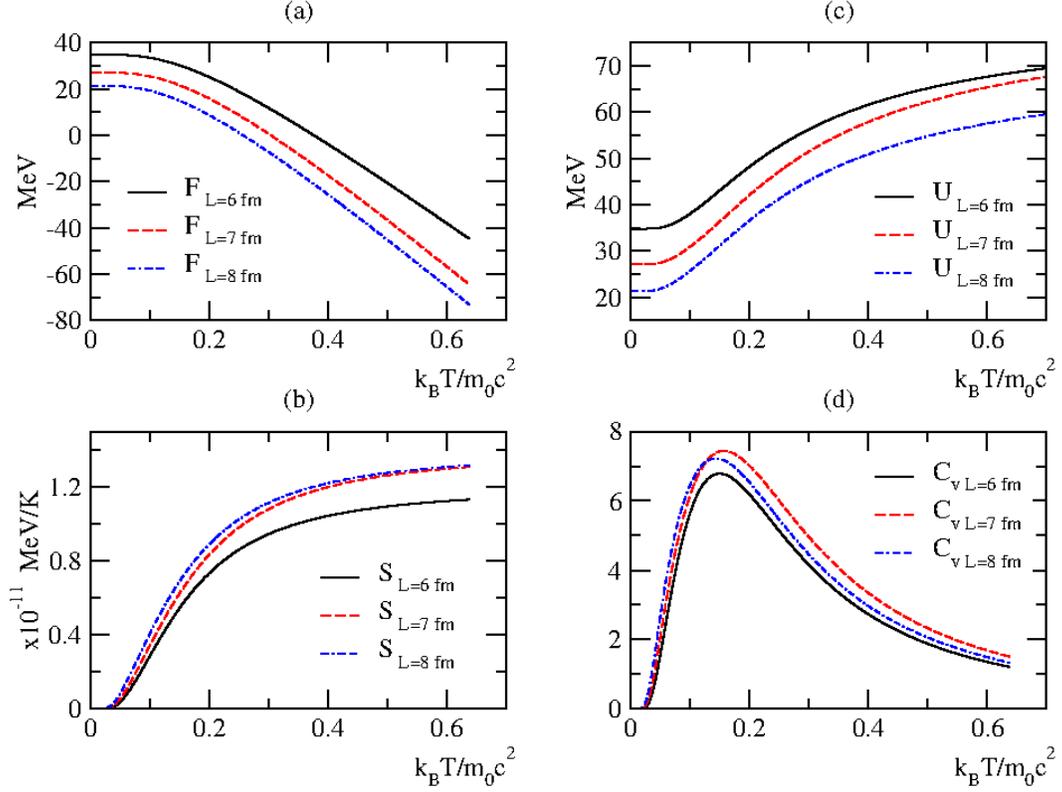}
   \caption{Reduced temperature versus the thermodynamic functions of a confined neutral pion particle in the attractive GSWSP energy wells that are constructed with the parameters given as follows:   $V_0=\frac{m_0c^2}{2}$, $W_0=-2m_0c^2$ and $a=1$ $fm^{-1}$.}  \label{fig17}
\end{figure}

\newpage
Next, we would like to investigate the effect of the increase of the effective length parameter on the strong attractive surface effect in a shallow potential energy well. We plot the GSWSP energy wells in fig. \ref{potfig04} (b). Note that, we tabulated the calculated energy eigenvalues in Table \ref{tab7}.  We illustrate the thermodynamic functions  in fig. \ref{fig17}.
The increase in effective distance increases the number of quantum states that can be entered for the second L value, but not the third L value. This effect is seen in Helmholtz free energy and entropy functions. On the other hand, specific heat functions exhibit almost the same behavior to a certain reduced temperature.

\begin{figure}[!htb]
\centering
\includegraphics[totalheight=0.45\textheight,clip=true]{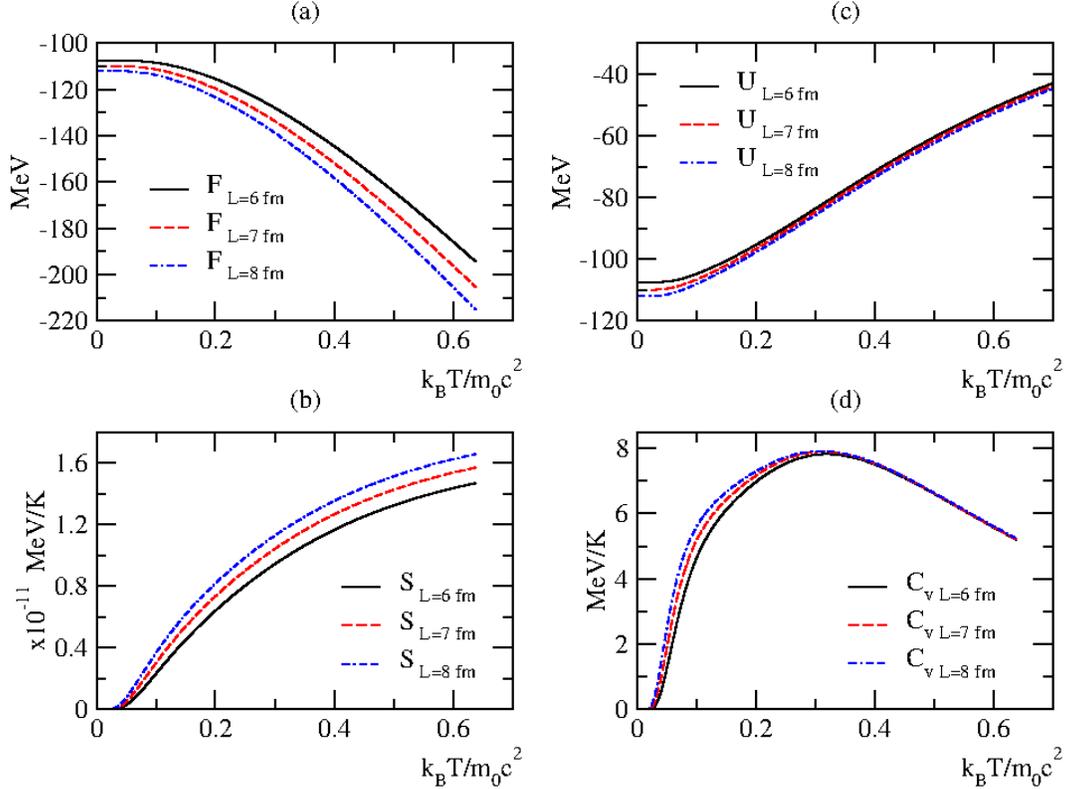}
   \caption{Reduced temperature versus the thermodynamic functions of a confined neutral pion particle in the attractive GSWSP energy wells that are constructed with the parameters given as follows:  $V_0=m_0c^2$, $W_0=-2m_0c^2$ and $a=1$ $fm^{-1}$.} \label{fig18}
\end{figure}

\newpage
Then, we would like to examine the effect of the increase of the effective length parameter on strong surface attractive effect in a deeper potential energy well. We plot the GSWSP energy wells in fig. \ref{potfig04} (d). Calculated energy eigenvalues are tabulated in Table \ref{tab8}. The obtained thermodynamic functions are demonstrated in fig. \ref{fig18}.

In this case, the potential wells are almost identical. The Helmholtz free energy slightly changes due to the increase of the available quantum states, thus the entropy function. Internal energy and specific heat functions are almost same.

We conclude that the increases  in the effective length in the deeper wells does not change the characteristic behavior of the thermodynamic functions.

\section{Conclusion}\label{sec:Concl}

In this paper, we present a comprehensive analysis of the thermodynamic functions of a confined neutral pion in the GSWSP energy wells. After brief overview of the known results, we discuss the GSWSP energy wells and the physical interpretation of the model. Note that in this paper, we study the shallow and / or deep wells that have strong surface interaction effects in addition to the volume effect. We illustrate comparative plots of the GSWSP wells and then, we give the bound state solution of the KG equation in the SS limit. We would like to emphasize that, this solution is obtained in our previous work.  We determine the neutral pion as the confined particle in the GSWSP energy wells. We calculate and tabulate the energy spectra in shallow and deep wells. After we introduce the partition function and the considered thermodynamic functions, we calculate them. We compare the Helmholtz free energy, entropy, internal energy, and specific heat functions of the neutral pion that is trapped in the GSWSP energy wells that have repulsive and attractive surface effects. We extend the analysis with including the changes in other potential shape parameters, namely the slope and effective length. Moreover, we discuss the changes in the thermodynamic functions, first in the repulsive case and then in the attractive case, by manipulating the potential shape parameters.

We conclude that in deep wells the increase of the slope parameter results with the overlap of the thermodynamic functions. Unlike the increase in the slope parameter, an increase in the effective length does not alter the character of thermodynamic functions in deep wells.

Although we deal with a toy potential well in this paper, we believe this analysis might help the specialist in the atomic and molecular field in real problems.

\section*{Acknowledgments}
One of the author, B.C.L.,  was partially supported for this work by the Turkish Science and Research Council \,\, (T\"{U}B\.{I}TAK) and Akdeniz University. The authors thank for the support given by the Internal Project of Excellent Research of the Faculty of Science of University Hradec Kr\'{a}lov\'{e}, "Studying of properties of confined quantum particle using Woods-Saxon potential".

\end{document}